\documentclass[aoas]{imsart}

\RequirePackage[T1]{fontenc}
\RequirePackage{xhfill}
\RequirePackage{amsthm,amsmath,amsfonts,amssymb,dsfont,relsize,bm}
\RequirePackage[authoryear]{natbib}
\RequirePackage[colorlinks,citecolor=blue,urlcolor=blue]{hyperref}
\RequirePackage{graphicx,soul}
\RequirePackage{caption, subcaption}
\RequirePackage{enumitem, hyphenat, float}
\RequirePackage{booktabs, longtable, array, lscape} 
\usepackage[usestackEOL]{stackengine}
\RequirePackage{tikz, pgf-pie, pgfplots}
\RequirePackage{tabularx, array}
\RequirePackage{verbatim}
\usepackage{siunitx}    

\DeclareMathOperator*{\argmin}{argmin}

\definecolor{babyblue}{rgb}{0.13, 0.67, 0.8}
\definecolor{junglegreen}{rgb}{0.16, 0.67, 0.53}

\newcolumntype{P}[1]{>{\centering\arraybackslash}p{#1}}
\newcolumntype{d}[3]{D{.}{.}{#1}}

\startlocaldefs
\newcolumntype{H}{>{\setbox0=\hbox\bgroup}c<{\egroup}@{}}
\newcommand{\ditto}[1][.4pt]{\xrfill[.6ex]{#1}~\textquotedbl~\xrfill[.6ex]{#1}}

\newcommand{\beginsupplement}{%
    \setcounter{table}{0}
    \renewcommand{\thetable}{S\arabic{table}}%
    \setcounter{figure}{0}
    \renewcommand{\thefigure}{S\arabic{figure}}%
    \setcounter{section}{0}
    \renewcommand{\thesection}{S\arabic{section}}%
}

\makeatletter
\renewcommand\listoffigures{%
        \@starttoc{lof}%
}
\makeatother

\theoremstyle{plain}

\newtheorem{algorithm}{Algorithm}

\theoremstyle{remark}

\endlocaldefs

\begin{document}

\begin{frontmatter}
\title{Model Determination for High-Dimensional Longitudinal Data with Missing Observations: An Application to Microfinance Data}
\thankstext{T1}{Corresponding author: Lotta Rüter, \href{mailto:lotta.rueter@kit.edu}{lotta.rueter@kit.edu}}

\begin{aug}
\author[A]{\fnms{Lotta} \snm{Rüter}\ead[label=e1]{lotta.rueter@kit.edu}} \and
\author[A,B]{\fnms{Melanie} \snm{Schienle}\ead[label=e2]{melanie.schienle@kit.edu}}

\address[A]{Institute of Statistics (STAT), Karlsruhe Institute of Technology (KIT)}
\address[B]{Heidelberg Institute of Theoretical Studies (HITS)}

\end{aug}

\begin{abstract}

We propose an adaption of the multiple imputation random lasso procedure tailored to longitudinal data with unobserved fixed effects which provides robust variable selection in the presence of complex missingness, high dimensionality and multicollinearity. We apply it to identify social and financial success factors of microfinance institutions (MFIs) in a data-driven way from a comprehensive, balanced, and global panel with 136 characteristics for 213 MFIs over a six-year period. 
We discover the importance of staff structure for MFI success and find that profitability is the most important determinant of financial success. Our results indicate that financial sustainability and breadth of outreach can be increased simultaneously while the relationship with depth of outreach is more mixed.


\end{abstract}


\end{frontmatter}
\section{Introduction}

Longitudinal data naturally emerge in many areas of research like biostatistics, sociology, health, labour and development economics. Such data are often incomplete, where a moderate share of overall missing values is distributed across observations so unfavourably that the amount of complete cases becomes negligible. Often, the underlying problem is also high-dimensional with a large cross-section dimension and few available observations, where covariates additionally might be highly correlated variables. Moreover, despite many included covariates in general, substantial amounts of subject-specific unobserved heterogeneity remain, that need to be captured by appropriate panel data methods. Aiming at model selection and inference, these challenges are usually addressed separately with the main approaches being the following. 
In practice, missing data is mostly either list-wise deleted, which can (obviously) lead to a substantial loss of valuable information, or imputed via multiple imputation. The latter method replaces the missing values with draws from probability distributions, commonly using either the joint posterior distribution of all variables with missing observations \citep{Little2019} or the conditional distribution of each variable condtioned on other variables in the data \citep{VanBuuren2007}. Many extensions have been proposed, e.g. to include interaction effects \citep{Goldstein2014}, general nonlinear effects \citep{Bartlett2015} or to account for sampling weights \citep{Zhou2016}. 
To deal with the high dimensionality of the data, variable selection methods such as the lasso and the elastic net can be used. These methods, however, have undesirable properties under multicollinearity of the data in that they tend to select only one of the highly correlated variables and shrink the impact of all others to zero \citep{Wang2011}.
Lastly, in the specific context of individual-level time-invariant heterogeneity, these methods may have ``poor estimation and inference properties'' \citep[see, e.g.][]{Belloni2016}.

In this paper, we jointly address the above points by building on the multiple imputation random lasso method introduced by \cite{Liu2016} and by adapting it to account for unobserved idioscyncratic effects in the data. We thereby obtain a methodology that yields robust results in the presence of high-dimensionality and multicollinearity for rather complex structures of missingness in longitudinal data as indicated above. The robustness we confirm by selection consistency across different base optimization citeria and sign consistency of effects for different ways of assessing post-selection effects. Our suggested aMIRL procedure generates values for missing entries via multiple imputation where the imputation step uses the panel structure of the observations and is as general as possible in imposed functional forms. In particular, we propose the use of a combination of mixed effects models and regression trees based on the random effects expectation maximisation (RE-EM) tree technique of \citep{Sela2012} for continuous variables and classification trees for binary covariates. The imputation step is followed by random lasso and stability selection which are both performed on the within-transformed imputed data for the inclusion of fixed effects in the final linear model.
The combination of imputation and stability selection enhanced model determination yields robust feature selection and model estimation in the presence of high-dimensionality and multicollinearity, see \cite{Wang2011} and \cite{Meinshausen2010}. An overview of further methods that perform variable selection on imputed data is provided by \cite{Zhao2017}.

We employ the aMIRL method for identifying success factors of microfinance institutions (MFIs) in a purely data-driven manner using the MIX Market data set from the World Bank Data Catalog that is characterised by many potentially relevant covariates with incomplete observations and therefore pronounced missingness.\footnote{\label{Data:Foot:MIX}\url{https://datacatalog.worldbank.org/dataset/mix-market}, retrieved on September 4, 2023.
} The problem is of substantial interest since over the last decades, microfinance institutions have been established to counter the problem that the poor have little access to financial help since they are not considered creditworthy by most banks due to their (obvious) lack of financial securities. MFIs hand out small credits (usually a few hundred USD) on terms and conditions different from those of common banks (e.g. \citealp{Morduch1999, Brau2004, VanRooyen2012, Quayes2015}). Instead of demanding financial securities, they rather come with obligations such as regular meetings with a liability group or participation in special training. Given that loans are accompanied by certain safety measures such as flexible repayment horizons and repayment limits, microcredits provide an important and successful tool to fight poverty not only on an individual but also on a macroeconomic level as shown by 
\cite{Yunus2009} and \cite{Imai2012}. Though the impact of microcredits is highly contextually sensitive \citep{Brau2004, Hulme2000}, multiple studies have determined significant individual-level effects that go beyond financial aid. These include the empowerment of women \citep{Cheston2002, Brau2004, Yunus2009}, the generation of businesses and new jobs \citep{Brau2004}, and positive changes in work ethics \citep{Banerjee2015}. Significant long-term effects on the welfare of villages and economies through higher wage and employment levels have been documented by 
\cite{Brau2004}, \cite{Imai2012} and \cite{Buera2021}. Given these positive effects of microcredits, it is argued that MFIs are most successful when they reach a large number (breadth of outreach) of especially poor (depth of outreach) borrowers while being financially sustainable and hence independent of external funding (e.g. \citealp{2007_Hartarska, Hermes2007, 2012_Bogan, Quayes2015}). 

In this work, we identify the determinants of such social and financial success in a data-driven way. To account for individual-level time-invariant heterogeneity via fixed effects, we construct a balanced panel comprising 1278 observations of 136 variables from 213 MFIs operating in 55 different countries that covers a span of six years (2009 to 2014). The final data set is thoroughly built ensuring that all potentially meaningful and important variables, as specified the literature (e.g. \citealp{2015_Basharat, Quayes2015, Hermes2019}), are contained, but redundant or uninformative measures are omitted. We apply the aMIRL method to account for the challenging structure and degree of missingness (95.5\% of the observations are incomplete with 13.7\% missing values in total), high-dimensionality and multicollinearity while including MFI-specific fixed effects.

To demonstrate the importance of accounting for the unobserved heterogeneity in the data as well as the additional robustness in the variable selection step, we compare our aMIRL results with fixed effects and pooled regression results from the original MIRL method as well as conventional lasso-estimates with column mean imputations. We further supplement the pooled regression results of the balanced panel data set with those of a large unbalanced data set with 3846 observations of 1026 MFIs located in 100 different countries from 2007 to 2018.

To our knowledge, this paper is the first to quantify and determine the importance of the personnel structure (rather than focusing on the role of management only as e.g. analysed in \cite{Kyereboah2008} or described in \cite{Hermes2019}) as a key driver for the social success of MFIs.
Amongst others, a greater borrower-staff ratio and an increased number of employees further an MFI's outreach. Both financial and social success benefit on average from reduced roles of management or board. Other drivers of an MFI's overall success are greater financial performance (main determinant of financial success) and lower costs. Breadth of outreach can be increased by setting certain new staff incentives and targeting specific (new) borrower groups. Depth of outreach is associated with higher charged interest rates and risk of default as also noted by \cite{Yunus2009}. We find that financial sustainability and breadth of outreach can go hand in hand, while the relationship of financial success and depth of outreach is less pronounced. Our results confirm the presence of mission drift, meaning that reaching more borrowers can lead to targeting wealthier borrowers and thus deviating from the original mission of serving the poorest \citep{Armendariz_2011}. On the other hand, issuing smaller loans can help to increase an MFI's breadth of outreach. Our aMIRL models significantly outperform the results of existing studies that also analyse MFI success based on the MIX Market data set in terms of goodness of fit measured by $R^2$. For example, our models for financial sustainability yield an $R^2$ that is 0.54 higher (0.88 vs. 0.34) than that of \cite{Quayes2015} who also uses balanced panel data and employs a fixed effects model to estimate the effects of potential drivers on operational sustainability.

Most studies on microfinance success that use non-experimental data employ linear panel models with pre-selected predictors, where the choice of regressors often depends on data availability and observations with missing values in selected regressors are simply dropped \citep{Ayayi2010, Quayes2015}. They analyse the sustainability and/or outreach of MFIs with regards to certain aspects such as competition \citep{Assefa2013}, poverty reduction \citep{Khandker2005}, profit orientation \citep{Roberts_2013}, governance \citep{2007_Hartarska, Kyereboah2008}, and capital structure \citep{2012_Bogan}. \cite{Hermes2019} provide a systematic review of literature on the determinants of social and financial performance of MFIs and state that "research on MFI performance is still in its infancy". The objective of this study is to contribute to the existing literature by examining both financial and social success drivers, thereby allowing for a direct comparison of the determinants of these dimensions of MFI success, while simultaneously accounting for the challenging missingness structure in the data, performing data-driven variable selection and addressing the high-dimensionality in the data.

Our empirical findings complement theoretical economic model-driven studies that an\-a\-lyse specific aspects as, e.g. the performance of MFIs in the presence of competition \citep{McIntosh2005}, potential deviation from their mission to reach the poor \citep{Armendariz_2011}, the diffusion of microcredits \citep{Banerjee2013} and their impact on whole economies \citep{Buera2021}.

Lastly, a large portion of the microfinance literature stems from randomised experiments where, mostly, ordinary linear models with control variables and/or treatment dummies are used \citep{Field2008, McIntosh2008, Swain2009, Field2013, Berge2015, Banerjee2015}. Due to the specificity of these setups in spatial design and research question, however, strong assumptions are required to derive general implications from such data.

This work is organised as follows. The next section provides details on the construction of the balanced panel data used. Section~\ref{Sec:Method} describes the method under investigation and Section~\ref{Sec:Results} presents our empirical results. Final conclusions and an outlook are given in Section~\ref{Sec:Conc}. 

We have made the software code for the proposed aMIRL technique and all benchmark model available in a GitHub repository \url{https://github.com/lottarueter/aMIRL}. For replicability of the empirical study, the repository also contains the raw data and the code for all pre-processing steps.

\section{Data}\label{Sec:Data}

The data used in this work originate from the MIX Market data set, which was made available in the World Bank Data Catalog on October 28, 2019.\footref{Data:Foot:MIX}
MIX is an acronym for Microfinance Information eXchange, Inc., a provider of global self-reported financial and social performance data of microfinance institutions launched in 2002 \citep{Imai2012}.  The MIX Market data set is the largest, most reliable data source on MFIs and comprises both financial and social performance data. In particular, our raw data is obtained from an inner join of the \textit{Financial Performance Data Set in USD} and the \textit{Social Performance Data Set} of the MIX Market data, i.e., it comprises all MFIs that appear in both of the source data sets. Concerning reliability of the data source, according to the World Bank's Data Catalog, data collection and reporting for MIX took place "in line with broadly recognised reporting standards within microfinance and inclusive finance".\footref{Data:Foot:MIX} That means, predetermined reporting formats were used and internal as well as external cross-checks were performed for validation \citep{Quayes2015}. For replicability, all following data pre-processing steps described below and in the online supplementary material are available as software code in the GitHub repository \url{https://github.com/lottarueter/aMIRL}. To ensure reproducibility, we also kept the (sub)categories and variable names of the original data set while constructing the panel used in this paper, see Table \ref{Appendix:Tab:variable_transformation} in the online supplementary material.\footnote{Note that the character "\textit{$>$}" functions as a subcategory indicator in the variable names of the original data set. Hence, the variable \textit{Av. Loan Size $>$ Gender $>$ Female} denotes the variable \textit{Av. Loan Size of Female Borrowers}. We employ a similar notation in this work. Instead of "\textit{$>$}" we use "\textit{$\triangleright$}".} However, for reasons of readability and interpretation, we introduce more compact names for the resulting variables and group them into different, new factors which are represented by different colours as shown in Figure~\ref{Data:Fig:Piechart-Factors}.

{\centering
\begin{figure}[htbp]
    \resizebox{12.7cm}{6.3cm}{
    \begin{footnotesize}
			\tikz{		
			\pie[sum = auto, color = {orange!20, orange!40, orange!60, babyblue!70, babyblue!40, babyblue!20, gray!40}, 
			hide number]
			{	14/\textsc{Organisational Strategy},
				18/\textsc{Personnel Structure},
				9/\textsc{Services},
                8/\textsc{Costs},
				6/\textsc{Financial Performance},
                13/\textsc{Financing Structure},
				68/\textsc{Operational Aspects}
			}
		}
		\end{footnotesize}
	}
    \caption[Proportion of variables per social, operational and financial factors in the final data set.]{Proportion of variables per social ({\color{orange} orange}), operational ({\color{gray} gray}) and financial ({\color{babyblue} blue}) factors in the final data set.}
    \label{Data:Fig:Piechart-Factors}
\end{figure}
}

Note that the raw data consists of a largely unbalanced panel comprising the years 2007 to 2018 where from year to year many new MFIs appear, existing ones dissappear, and sometimes previously contained ones reappear (see Table \ref{Data:Tab:DatAvailSummary}). Conditional on even the largest set of observable characteristics for MFIs, however, it is well known that MFIs across the globe are quite heterogeneous and have their own peculiar specificity (see, e.g. \cite{Fall2023}). In order to capture such MFI-specific heterogeneity in a practically feasible fixed effects approach, we construct a balanced panel from the unbalanced raw data. This is the main basis for our empirical analysis. If we used unbalanced data instead, different sample sizes in the time dimension $T=T_i$ would affect the within-transformation differently for different MFIs resulting in non-standard statistical properties of the final estimates that are beyond the scope of the present paper. We would further still have to the exclude the many cases with $T_i<2$ that do not allow for the identification of a fixed effect.

In the construction of the balanced panel, we select  the considered timing and time span with consecutively available MFIs such that it yields the maximum number of available balanced panel elements. An MFI is classified as ``available'' in the raw data if for all time points in the respective period it has at least one non-zero entry per target variable and across all factors specified below in detail. Please see Table \ref{Data:Tab:DatAvailSummary} that displays all possible allocation options of a balanced panel within the unbalanced raw one with corresponding sample sizes. As the optimal window for our analysis, we chose $w^*=[2009,2014]$ with $N_{w^*}\times T_{w^*}= 213 \times 6 = 1278$ available panel observations. We prefer this to the case that only comprises four years instead of six with an only slightly larger sample size of $1284$. For our goal of determining the success factors of MFIs, we consider a larger time horizon crucial for capturing the evolution of MFIs but also for improving the precision of fixed effect estimates and the respective within-transformations in a panel model. 

\begin{table}[htbp]
\begin{footnotesize}
    \caption{Data availability for balanced panel allocations within the raw unbalanced data}
	\begin{tabularx}{\textwidth}{lHccccccccccc}
		\toprule
		\textbf{} & $\textbf{2007}$ & $\textbf{2008}$ & $\textbf{2009}$ & $\textbf{2010}$ & $\textbf{2011}$ & $\textbf{2012}$ & $\textbf{2013}$ & $\textbf{2014}$ & $\textbf{2015}$ & $\textbf{2016}$ & $\textbf{2017}$ & $\textbf{2018}$ \\
		\midrule
		\textbf{2008} & & \Centerstack{$199$ \\ $(199)$} & \Centerstack{$280$ \\ $(140)$} & \Centerstack{$417$ \\ $(139)$} & \Centerstack{$540$ \\ $(135)$} & \Centerstack{$625$ \\ $(125)$} & \Centerstack{$666$ \\ $(111)$} & \Centerstack{$700$ \\ $(100)$} & \Centerstack{$56$ \\ $(7)$} & \Centerstack{$18$ \\ $(2)$} & \Centerstack{$0$ \\ $(0)$} & \Centerstack{$0$ \\ $(0)$} \\ \specialrule{0.25pt}{0.5\jot}{0.5\jot}
		\textbf{2009} & & & \Centerstack{$330$ \\ $(330)$} & \Centerstack{$636$ \\$(318)$} & \Centerstack{$924$ \\ $(308)$} & \Centerstack{$1084$ \\ $(271)$} & \Centerstack{$1160$ \\ $(232)$} & \Centerstack{\textbf{1278} \\ \textbf{(213)}} & \Centerstack{$63$ \\ $(9)$} & \Centerstack{$24$ \\ $(3)$} & \Centerstack{$9$ \\ $(1)$} & \Centerstack{$10$ \\ $(1)$} \\ \specialrule{0.25pt}{0.5\jot}{0.5\jot}
  		\textbf{2010} & & & & \Centerstack{$375$ \\ $(375)$} & \Centerstack{$728$ \\$(364)$} & \Centerstack{$912$ \\ $(304)$} & \Centerstack{$1012$ \\ $(253)$} & \Centerstack{$1140$ \\ $(228)$} & \Centerstack{$60$ \\ $(10)$} & \Centerstack{$28$ \\ $(4)$} & \Centerstack{$16$ \\ $(2)$} & \Centerstack{$18$ \\ $(2)$} \\ \specialrule{0.25pt}{0.5\jot}{0.5\jot}
        \textbf{2011} & & & & & \Centerstack{$552$ \\ $(552)$} & \Centerstack{$918$ \\$(459)$} & \Centerstack{$1101$ \\ $(367)$} & \Centerstack{$1284$ \\ $(321)$} & \Centerstack{$55$ \\ $(11)$} & \Centerstack{$30$ \\ $(5)$} & \Centerstack{$21$ \\ $(3)$} & \Centerstack{$16$ \\ $(2)$} \\ \specialrule{0.25pt}{0.5\jot}{0.5\jot}
        \textbf{2012} & & & & & & \Centerstack{$612$ \\ $(612)$} & \Centerstack{$922$ \\$(461)$} & \Centerstack{$1188$ \\ $(396)$} & \Centerstack{$76$ \\ $(19)$} & \Centerstack{$35$ \\ $(7)$} & \Centerstack{$18$ \\ $(3)$} & \Centerstack{$14$ \\ $(2)$} \\ \specialrule{0.25pt}{0.5\jot}{0.5\jot}
        \textbf{2013} & & & & & & & \Centerstack{$617$ \\ $(617)$} & \Centerstack{$1042$ \\$(521)$} & \Centerstack{$84$ \\ $(28)$} & \Centerstack{$44$ \\ $(11)$} & \Centerstack{$35$ \\ $(7)$} & \Centerstack{$12$ \\ $(2)$} \\ \specialrule{0.25pt}{0.5\jot}{0.5\jot}
        \textbf{2014} & & & & & & & & \Centerstack{$671$ \\ $(671)$} & \Centerstack{$82$ \\$(41)$} & \Centerstack{$51$ \\ $(17)$} & \Centerstack{$32$ \\ $(8)$} & \Centerstack{$15$ \\ $(3)$} \\ \specialrule{0.25pt}{0.5\jot}{0.5\jot}
        \textbf{2015} & & & & & & & & & \Centerstack{$46$ \\ $(46)$} & \Centerstack{$34$ \\$(17)$} & \Centerstack{$24$ \\ $(8)$} & \Centerstack{$12$ \\ $(3)$} \\ \specialrule{0.25pt}{0.5\jot}{0.5\jot}
        \textbf{2016} & & & & & & & & & & \Centerstack{$155$ \\ $(155)$} & \Centerstack{$146$ \\$(73)$} & \Centerstack{$123$ \\ $(41)$} \\ \specialrule{0.25pt}{0.5\jot}{0.5\jot}
		\textbf{2017} & & & & & & & & & & & \Centerstack{$111$\\$(111)$} & \Centerstack{$110$ \\$(55)$} \\ \specialrule{0.25pt}{0.5\jot}{0.5\jot}
		\textbf{2018} & & & & & & & & & & & & \Centerstack{$163$ \\ $(163)$} \\
		\bottomrule
		\multicolumn{13}{l}{%
        \begin{minipage}{0.96\linewidth} \vspace{0.3cm}
            \linespread{1}
            \footnotesize \textbf{Note:} The rows indicate the start year of the window, the columns the end year of the considered balanced panel allocation. In each cell, the top number marks the size of the respective balanced panel $N_w \times T_w$, where $N_w$ in parenthesis below is the number of MFIs with at least one non-zero entry in each year per given window in the raw data set and $T_w$ denotes the corresponding window length. We do not display windows with the start year 2007, as only 15 MFIs were available in this year in our raw data set.
        \end{minipage}%
		}
	\end{tabularx}%
	\label{Data:Tab:DatAvailSummary}%
 \end{footnotesize}
\end{table}

This results in a balanced panel data set that contains observations of 213 MFIs operating in 55 different countries for six consecutive years (2009 to 2014). For all MFIs in the balanced panel, we have observations on the considered three different target variables and 136 explanatory factors as detailed in Subsections \ref{Data:Sec:TarVars} and \ref{Data:Sec:factors} below. Detailed summary statistics are provided in Table \ref{Appendix:Tab:DescrStatsBalanced} in the online supplementary material. The data structure in the final balanced panel contains major challenges that comprise in particular the degree and structure of missingness in covariates and that require specialised statistical techniques to address them. Please see Subsection~\ref{Data:Sec:FinData} below. 

Due to the limited size sample size of the constructed balanced panel, we use an unbalanced panel that comprises all years 2007 to 2018 for robustness checks (see Section~\ref{Results:Sec:RobChecks}). It comprises 1026 MFIs and a number of 3846 observations for the same set of target variables and covariates as in the balanced panel. Its summary statistics are given in Table \ref{Appendix:Tab:DescrStatsUnbalanced}. The table illustrates that not only the non-consecutive years in the panel but also the structure and degree of missingness of covariates poses an additional challenge in this case .

\subsection{Choice of the Dependent Variables} \label{Data:Sec:TarVars}
Since poverty reduction should be the main objective of MFIs, they are considered successful if they maximise their social impact rather than, e.g, their profits.\footnote{Profit maximisation is argued to be a dangerous goal by \cite{Yunus2009} since it likely leads to the exploitation of the poor rather than their benefit.} Given the various positive effects of microcredits mentioned in the introduction, MFIs are defined as socially successful if they reach (i) many and (ii) particularly poor borrowers (e.g. \citealp{Cull2007}). The first dimension, breadth of outreach, is measured by the logarithm of the \textit{Number of Active Borrowers}, \textit{log(NAB)} \citep{2007_Hartarska, Quayes2015}. However, direct measures of the second dimension, depth of outreach, such as the income level of borrowers, are not available. Under the assumption that poorer borrowers generally receive smaller loans, we use the negative of the logarithm of \textit{Average Loan Balance per Borrower / GNI per Capita}, \textit{--log(ALBG)}, as a standardised proxy as done by \cite{2007_Hartarska} and \cite{Armendariz_2011}, amongst others. To maximise long-term social success, MFIs are required to be operationally sustainable and hence independent of external funding. The money originally used for financing can then be used to set up new MFIs, which increases the overall social impact. An MFI is operationally sustainable when its revenues exceed its costs. The most common indicator of financial MFI performance is therefore \textit{Operational Self-Sufficiency}, \textit{OSS} \citep{2007_Hartarska, Assefa2013, Quayes2015}. It relates the total income of the MFI to the expenditure required for its operation, see Table \ref{Appendix:Tab:variable_transformation} for the exact definition. 
To summarise, we study success determinants of MFIs with regards to \textit{log(NAB)}, \textit{--log(ALBG)} and \textit{OSS}.

\subsection{Included Explanatory Factors} \label{Data:Sec:factors}
We only include variables with less than 50\% missing entries and perform further transformations which are described in detail in Section \ref{Data:Sec:PreProcSteps} and Table \ref{Appendix:Tab:variable_transformation} in the online supplementary material. For the resulting full list of the 136 considered variables and details of their definitions as well as their respective transformations we also refer to Section~\ref{Data:Sec:PreProcSteps} and Table \ref{Appendix:Tab:variable_transformation} in the online supplementary material. The final set of characteristics in the balanced panel covers a wide span of potential success factors ranging from performance indicators and measures of the financial structure of the MFI to its borrower structure, aggregated personnel data, information on staff incentives and development targets.

The 105 included variables from the \textit{Financial Performance Data Set in USD} render information on the distribution of the loan portfolio, the client as well as the personnel structure and the financial situation of each MFI. Here, we focus on an overview of the considered different categories in the original data set. Data on the shares of the loan portfolio and loan sizes per borrower type, lending methodology and credit delay are given in the categories \textit{Clients}, \textit{Credit Products}, \textit{Delinquency} and \textit{Outreach}. Additionally, information on the number of (new) borrowers and the number of loans outstanding is included. Further, balance sheet positions and ratios describing the financing, risk and liquidity structures as well as the income and expenses are presented in the corresponding categories \textit{Balance Sheet}, \textit{Financing Structure}, \textit{Risk \& Liquidity} and \textit{Income}. Financial performance, productivity and efficiency measures and further revenue and expenses sizes are given by the categories \textit{Financial Performance}, \textit{Productivity \& Efficiency} and \textit{Revenue \& Expenses}. The personnel structure is portrayed in the categories \textit{Infrastructure} and \textit{Social Performance}. We implicitly account for macroeconomic characteristics, since the variables \textit{GNI per Capita} and \textit{Inflation Rate} are contained in quantities such as \textit{Av. Loan Bal. $/$ GNI p. c.}, \textit{Av. Salary $/$ GNI p. c.} and \textit{Yield on GLP (Real)}.\footnote{The explicit consideration of these variables only increases the multicollinearity in the data, but does not lead to significantly different results. Corresponding results can be made available on request.}

The 31 considered variables from the \textit{Social Performance Data Set} contain information on client protection measures, staff incentives and additional services provided by the MFI, see the categories \textit{Client Protection}, \textit{Governance \& HR} and  \textit{Products \& Services}. The category \textit{Social Goals} further comprises development goals, a quantification of the MFI's focus on poverty reduction, and its target markets.

\subsection{Characteristics and Challenges of the Final Data Set} \label{Data:Sec:FinData}
\begin{figure}[!h]
	\centering	\vspace{-1cm}
	\includegraphics[width=\textwidth]{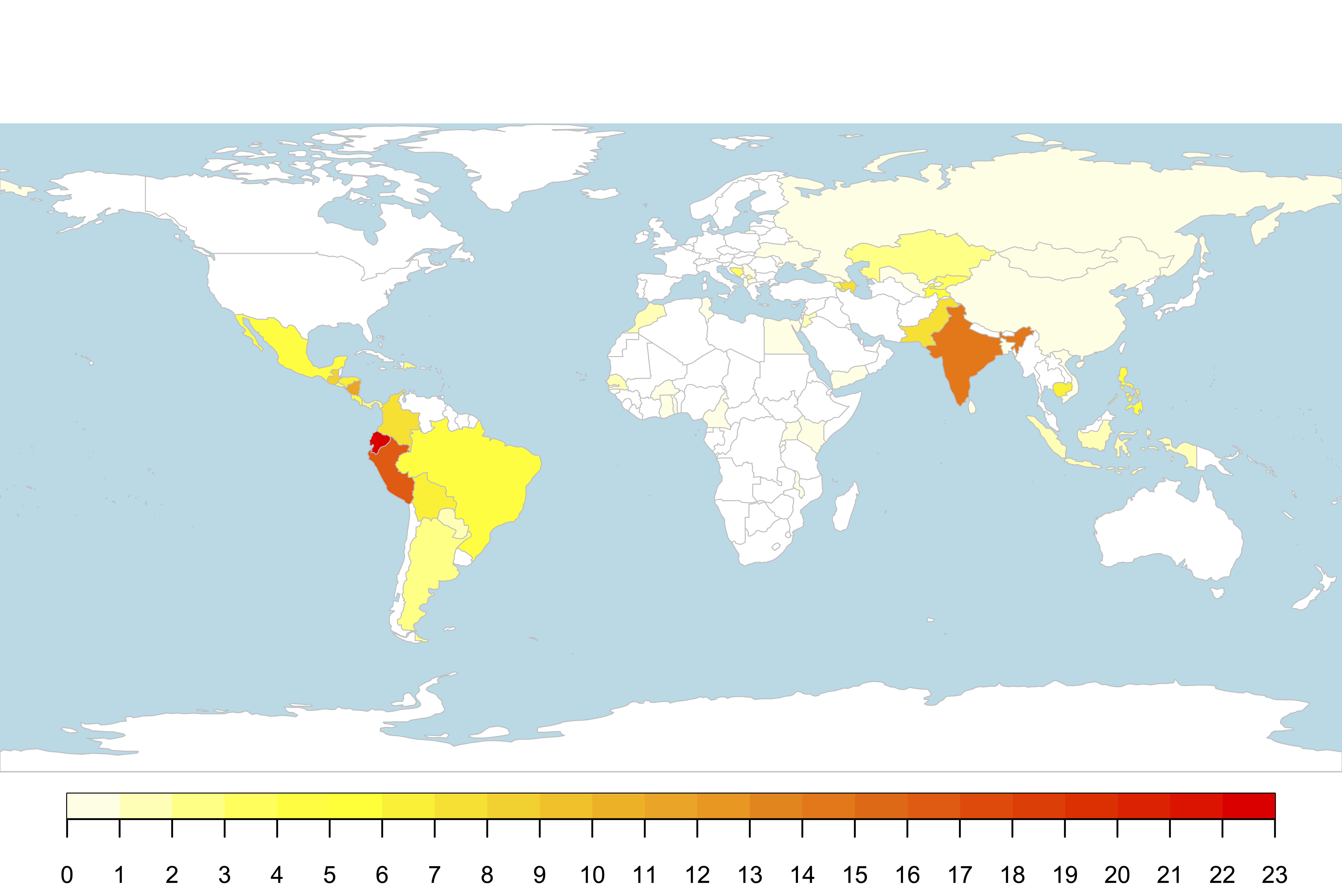}
	\caption[Map with the number of MFIs per country in the final data set.]{Map with the number of MFIs per country in the final data set.}
	\label{Data:Fig:CountriesMap_blue}
\end{figure}

53\% of the included MFIs operate in South America and 35\% in Asia while Europe and Africa each provide 6\% of the data, as shown on the map in Figure~\ref{Data:Fig:CountriesMap_blue} and Table \ref{Data:Tab:RegionsContinentsMFIs}. More than 75\% of the MFIs are operationally sustainable (i.e. \textit{Operational Self-Sufficiency} $\ge1$). 50\% have assets of more than USD 25 million and a gross loan portfolio of more than USD 20 million while serving at least 25,813 borrowers, see Table \ref{Appendix:Tab:DescrStatsBalanced} in the online supplementary material. In addition, clearly only MFIs that have been operating successfully for at least six years are included. These aspects should be taken into account when interpreting our results. While our findings may contain practical insights for the establishment of new MFIs globally, statistically we can only identify variables that are associated with increased success of already established MFIs operating primarily in South America and Asia.

\begin{table}[htbp]
\begin{footnotesize}
	\centering	
	\caption[Number of different MFIs (countries) per continent and region in the final data set.]{Number of different MFIs (countries) per continent and region in the final data set.}
	\begin{tabularx}{0.9\textwidth}{p{4cm}P{1.25cm}P{1.25cm}P{1.25cm}P{1.25cm} | P{1.25cm}}\toprule
		                            & \textbf{Africa} & \textbf{America} & \textbf{Asia} & \textbf{Europe} & \\\midrule
		\textbf{East Asia \& Pacific} 			& $0$      & $0$        & $18\;(7)$ & $0 $       & $18\;(7)$\\
		\textbf{Europe \& Central Asia} 		& $0$      & $0$        & $27\;(7)$ & $12\;(7)$  & $39\;(14)$ \\
		\textbf{Latin America \& Caribbean} 	& $0$      & $113\;(15)$& $0$       & $0$        & $113\;(15)$\\
		\textbf{Middle East \& North Africa}    & $4\;(3)$ & $0$        & $5\;(4)$  & $0$        & $9\;(7)$\\
		\textbf{South Asia}						& $0$      & $0$        & $25\;(4)$ & $0$        & $25\;(4)$\\
		\textbf{Sub-Saharan Africa} 			& $9\;(8)$ & $0$        & $0$       & $0$        & $9\;(8)$\\\midrule
		& $13\;(11)$& $113\;(15)$& $75\;(22)$& $12\;(7)$ & $213\;(55)$\\ \bottomrule
	\end{tabularx}%
	\label{Data:Tab:RegionsContinentsMFIs}
\end{footnotesize}
\end{table}

Many of the potential regressor variables are moderately or highly correlated, see, e.g. Figure~\ref{Results:Fig:OSS_corr_sd} in Section~\ref{Results:Sec:RobChecks}.
Depending on the target variable, we remove those regressors that are by definition almost identical to the target in the respective analysis.\footnote{In particular, for target variable \textit{log(NAB)} we exclude \textit{\# Active Borrowers} and \textit{\# Loans Outstanding} from the set of potential regressors. For \textit{--log(ALBG)} we remove \textit{Av. Loan Bal. $/$ GNI p. c.} and \textit{Av. Outst. Bal. $/$ GNI p. c.}. \textit{log(NAB)} and \textit{--log(ALBG)}, are only used as target variables. In the set of potential regressors, their untransformed versions \textit{\# Active Borrowers} and \textit{Av. Loan Bal. $/$ GNI p.c.} are included.} The resulting maximum correlation between an outcome variable and a potential regressor is 0.73. 

The main statistical challenge of the final balanced panel data set, however, consists in the substantial amount of missingness in covariates and target variables as documented in  Table \ref{Appendix:Tab:DescrStatsBalanced} in the online supplementary material. In particular, only for 58 of all considered 1278 instances $it$ ($i = 1,...,213$ and $t=1,...,6$ with $213\times 6 = 1278$) in the balanced panel, all 136 factors are completely observed.\footnote{In the unbalanced panel from 2007 to 2018 in the comparison study, only 177 of 3846 instances are complete. See Table \ref{Appendix:Tab:DescrStatsUnbalanced} for details in this case.} Note that it is the allocation of the missing values that causes 95.5\% instances in the panel to suffer from incompletely observed characteristics. When counting the aggregate number of unobserved factors and target variables across all MFIs and time points relative to the total possible number of $1278\times 136$ in the balanced panel there are only 13.7\% missing values, causing the degree of overall missingness to appear rather moderate. Concerning the allocation of missing observations across covariates, note that there is no inherent ordering in the factors and no further structure in missing observations across covariates. This implies that the missingness is nonmonotone in covariates, i.e., variables of the predictor set cannot be sorted in such a way that the missingness of one variable implies the missingness of all subsequent variables. 

The multicollinearity and missingness in the data hence pose non-trivial challenges for the model determination and estimation in this work, which requires specialised statistical techniques.

\section{Method}\label{Sec:Method}
Our goal to identify MFI success factors in a purely data-driven manner using the panel data presented calls for a method with corresponding characteristics. First, the method must be able to identify the most important variables and filter out their singular effects in the presence of high-dimensionality and correlated variables. Second, it must be able to handle the considerable amount of complex missingness. Third, it should incorporate the longitudinal structure of the data.

We build on the Multiple Imputation Random Lasso  procedure (MIRL) as introduced by \cite{Liu2016}, which combines random lasso for variable determination with multiple imputation and stability selection, specifically to handle the high degree of missingness in the data. Random lasso is a two-step procedure which provides variable selection and parameter estimation of linear models in high-dimensional settings with multicollinearity. It is shown to outperform similar existing methods in such cases in terms of prediction accuracy \citep{Wang2011}. Combined with multiple imputation, it systematically accounts for missingness in the data \citep{Azur2011}. The incorporation of stability selection further yields an importance ranking of all variables where a data-adapted threshold determines the final set of predictors. In particular, the procedure with stability selection requires much weaker conditions than the original lasso in order to still yield consistent variable selection for dependent data  \citep{Meinshausen2010}.

We suggest an adaptation of the original MIRL approach that accounts for the longitudinal structure in our data. For the estimation and variable selection step, we propose a linear fixed effect panel setting, where MFI-specific unobserved characteristics are captured by time-invariant individual effects $\alpha_i$, so that
\begin{equation} \label{meth:eq:fem}
	y_{it} = \alpha_{i} + \mathbf{x}_{it}^\prime \pmb{\beta} + \epsilon_{it},
\end{equation}
with $y_{it}$ denoting the value of the target variable for MFI $i$ at time $t$ for $i=1,\ldots, N$ and $t=1, \ldots, T$, where $N=213$ and $T=6$ in our case. Moreover, $\mathbf{x}_{it} = (x_{1, it}, ..., x_{p, it})^\prime$ comprises the observations of $p$ explanatory variables for $i$ at time $t$, $\pmb{\beta}= (\beta_1, ..., \beta_p)^\prime$ contains the linear effects of $x_1, ..., x_p$ on $y$, and $\epsilon_{it}$ is the random error term of the model for individual $i$ at time $t$ satisfying the standard panel fixed effects exogeneity assumptions. Using the usual within-transformation as done by \cite{Belloni2016}, a pilot estimate $\pmb{\hat{\beta}}$ of the full model can be obtained via OLS on the time-demeaned model, i.e.
\begin{align} \label{Eq:Results:within1}
	\ddot{y}_{it} & = \ddot{\mathbf{x}}_{it}^\prime \pmb{\beta} + \ddot{\epsilon}_{it}
\end{align}
with $\ddot{y}_{it} =  y_{it}-\bar{y}_i$ where  $\bar{y}_i=\frac{1}{T}\sum_{t=1}^Ty_{it}$ and $\ddot{\mathbf{x}}_{it}$ and $\ddot{\epsilon}_{it}$ are transformed respectively, see \cite{Wooldridge2012}. We take this time-demeaned form~\eqref{Eq:Results:within1} as the starting point for Steps 2--4 of our adapted MIRL procedure that addresses the complex challenges in the data such as high-dimensionality, missingness and multicollinearity.

\subsection{The Adapted MIRL Procedure}
The adapted MIRL (aMIRL) technique comprises four steps which are presented in detail in the following. In essence, the aMIRL simultaneously selects variables and estimates parameters across panel bootstrap samples of a sequence of imputed data sets, where the final parameter estimates are robust aggregates across samples and relevant components are ranked and selected  according to stability selection \citep{Meinshausen2010}. Thus, the first step of aMIRL generates multiple imputations and carries out the standardisation and the within-transformation of the data. Steps 2 and 3 represent an adaption of the random lasso algorithm \citep{Wang2011, Liu2016} in the panel stetting and Step 4 comprises the selection and ranking of variables performed by stability selection.

\subsection*{Step 1: Imputation and Standardisation}
\label{Method:Sec:Step1}
All missing values in the data are imputed using the Multivariate Imputation by Chained Equations (MICE) technique \citep{Raghunathan2001, VanBuuren2007, Azur2011}. It is based on the Missing At Random assumption, where the probability that a value is missing depends only on the observed data, but not on unobserved components \citep{Schafer2002}. The MICE procedure yields $M$ completed data sets that differ in their final imputed values. We choose $M=10$ in line with the literature that suggests a range from $M=5$ to $M=40$ but offers no further criterion for the determination of $M$ \citep{Azur2011}.  For each $m \in \{1,...,M\}$ we obtain the final imputed values according to the following procedure.

 We start by replacing each missing value in the data with the mean of the available observations of the corresponding variable (\textit{place holder imputation}). We then fix a certain variable $x_j$ and set its imputed values back to missing. The observed values of $x_j$ are then regressed on the imputed and observed values of all other variables in the data set. We use nonlinear regression and classification trees to countervail the complexity and correlation challenges in the data. For continuous $x_j$, we employ RE-EM trees by \cite{Sela2012} which are regression trees with random effects that accommodate nonlinearities in the data while taking the panel structure into account. The resulting imputation procedure for continuous outcomes is described in Algorithm \ref{Alg:REEMTree}.
 
\begin{algorithm}{Imputation of Continuous Covariate $x_j$ Using the Panel Structure.}
\label{Alg:REEMTree} 
\vspace{0.2cm}

 Variable $x_{j}$ is modelled as $x_{j,it} = u_i + f(\mathbf{x}_{-j,it}) + e_{it}$, where $u_i$ is an MFI-specific random effect, $\mathbf{x}_{-j,it}$ contains the values of all $p$ regressors except $x_j$ for MFI $i$ at time $t$, $f$ is an unknown function that will be approximated via regression trees and $e_{it}$ denotes the error term. Further, we denote by $\mathcal{I}_j^\text{obs}$ the set of indices $(i,t)\in \{(1,1),...,(213,6)\}$, for which $x_{j}$ is observed and by $\mathcal{I}_j^\text{miss}$ the set of indices $(i',t')$ for which the values of $x_j$ are missing.
 
 \vspace{0.2cm}
\noindent \textbf{Part A (Estimation)}: Estimate the RE-EM trees using all observations $(i,t)\in \mathcal{I}_j^\text{obs}$:
\begin{enumerate}
    \item \textbf{Initialization}: Begin by setting the initial estimated random effects $\hat{u}_i$ to zero.

    \item \textbf{Iterative Estimation}: Repeat the following steps until the estimated random effects $\hat{u}_i$ converge, i.e. until the change in the restricted likelihood function falls below a pre-defined tolerance level.
    \begin{enumerate}
        \item \textbf{Tree Estimation}: Estimate a regression tree that approximates the function $f$, using the adjusted target variable $x_{j,it} - \hat{u}_i$ and the attributes $\mathbf{x}_{-j,it} = (x_{1,it}, \ldots, x_{j-1,it},$ $ x_{j+1,it}, \ldots, x_{p,it})'$. The vector of attributes $\mathbf{x}_{-j,it}$ contains the observed or latest imputed values of all remaining variables. From this regression tree, generate a set of indicator variables $I(\mathbf{x}_{-j,it} \in g_p)$, where $g_p$ represents the terminal nodes of the tree.

        \item \textbf{Mixed Effects Model Fit}: Fit a linear mixed effects model of the form
        \[
        x_{j,it} = u_i + I(\mathbf{x}_{-j,it} \in g_p) \, \mu_p + e_{it},
        \]
        where $\mu_p$ is the prediction for leaf $p$. From the fitted model, extract the updated estimates $\hat{u}_i$ for the random effects.
    \end{enumerate}

    \item \textbf{Response Adjustment}: After convergence, replace the predicted response at each terminal node of the tree with the estimated $\hat{\mu}_p$ derived from the linear mixed effects model in 2. b).
\end{enumerate}

\noindent \textbf{Part B (Imputation)}: Replace the missing values of $x_j$ for $(i',t')\in \mathcal{I}_j^\text{miss}$ as follows:

\begin{itemize}
    \item \textbf{Case 1}: MFI $i'$ is contained in the estimation sample, i.e. $\exists \, t:(i',t)\in \mathcal{I}_j^\text{obs}$. We impute $x_{j,i't'} := \hat{u}_{i'} + I(\mathbf{x}_{-j,i't'} \in g_p) \, \hat{\mu}_p$.

    \item \textbf{Case 2}: MFI $i'$ is not contained in the estimation sample, i.e. $\nexists \, t:(i',t)\in \mathcal{I}_j^\text{obs}$. We have no basis for estimating the random effect $u_{i'}$ and therefore set $u_{i'}=0$. We impute $x_{j,i't'}:=I(\mathbf{x}_{-j,i't'} \in g_p) \, \hat{\mu}_p$.
    \end{itemize}
\end{algorithm}
 
The use of fixed effects instead of random effects is impracticable here, as the iterative nature of the estimation procedure would lead to an overestimation of the individual effects and an underestimation of the component $f$. This would particularly affect the imputation of missing values of MFIs $i'$ that are not included in the estimation data set (Case 2 in Part B of Algorithm \ref{Alg:REEMTree}). In this case, $u_{i'}$ cannot be estimated and is therefore set to zero, which would lead to imputed values close to zero. In the binary setting, the inclusion of random effects is not straightforward and we therefore rely on conventional classification trees for imputing the few dummy variables contained in the data. Subsequently, the missing values of $x_j$ are replaced with predictions from the fitted regression model, i.e. either using Algorithm \ref{Alg:REEMTree} or estimated classification trees in the case of binary outcomes.\footnote{For very few (0.65\%) of the missing observations, the RE-EM trees predicted values $<0$ or $>1$, although only values from $0$ to $1$ had been observed in the respective variable. We manually replaced these values with $0$ and $1$.} These imputed values then replace the corresponding placeholders or previously imputed values. Iterating through all of the $p$ variables and repeating the previously mentioned steps forms a \textit{cycle}. At the end of the first cycle, all missing values have been imputed once. In total, we perform $C$ cycles, updating the imputations in each cycle and permuting the order of the components $j$ in the updating steps. We set $C=20$, which is the maximum in the suggested range of 10 to 20 (see \citealp{VanBuuren2011a}) to cope with the strong correlations in our data \citep{White2011}. For each completed data set $m$, the ensuing lasso procedure requires coefficients of comparable size. The resulting data is therefore standardised to have zero mean and unit variance. In this fixed effects panel adaption of the algorithm, the data is further time-demeaned since we estimate $\pmb{\beta}$ using the within-transformed model~\eqref{Eq:Results:within1}.

\subsection*{Step 2: Bootstrap Samples and Importance Measures}
We obtain the parameter estimates by applying a slightly modified version of the random lasso, introduced by \cite{Wang2011}, which re-estimates a lasso model on multiple bootstrapped data sets. It has been shown to outperform the elastic-net method in terms of its efficiency in selecting or removing highly correlated variables and its flexibility in coefficient estimation.

From each of the $M$ imputed data sets, $B$ bootstrap samples are drawn, resulting in ${M \times B}$ bootstrapped data sets. To maintain the panel structure in each sample, we draw with replacement from vectors of $T=6$ observations per $N=213$ different MFIs. Based on the recommendations of \cite{Wang2011}, we choose $B=100$ and obtain $M\times B=1000$ bootstrapped data sets that are of the same size as the original $N\times T$ panel data set.

The subsequently computed importance measures are required for the subset selection of the random lasso procedure in Step 3. Intuitively, important (unimportant) variables $j$ are likely to have consistently large (small) lasso estimates $\hat{\beta}_j$ in different bootstrap samples, where the fixed effects lasso estimator $\pmb{\hat{\beta}}$ for model~\eqref{meth:eq:fem} solves the penalised regression
\begin{equation}\label{Eq:lassoest}
	\underset{\mathbf{\theta} \in \mathbb{R}^{^p}}{\min} \Vert \mathbf{\ddot{y}}-\mathbf{\ddot{X}}\mathbf{\theta} \Vert_{NT}^2 + \lambda^\text{OC} \Vert \mathbf{\theta} \Vert_1,
\end{equation}
where $\Vert z \Vert_{NT}^2= \dfrac{1}{NT}\sum_{i=1}^{N}\sum_{t=1}^{T} z_{it}^2$ is the empirical norm, $\Vert \mathbf{\theta} \Vert_1 = \sum_{j=1}^{p}\vert \theta_{j}\vert$ denotes the $\ell^1$-norm and $\mathbf{\ddot{X}}=(\mathbf{\ddot{x}}_{11}^\prime, \dots, \mathbf{\ddot{x}}_{1T}^\prime, \mathbf{\ddot{x}}_{21}^\prime, \dots , \mathbf{\ddot{x}}_{NT}^\prime)^\prime$ contains all $N\times T$ observations of the $p$ regressor variables. The tuning parameter $\lambda^\text{OC}\in \mathbb{R}$ is chosen in a data-driven way as outlined below in Section~\ref{Method:Sec:OC}, minimising one of the three optimality criteria (OC) BIC, AIC and Mallows' C$_p$. We obtain lasso pre-estimates $\pmb{\hat{\beta}}^{(b), \text{OC}}_m$ for every pair of imputation and bootstrap sample $(m,b)$ and consider a variable $j$ as relevant if $\hat{\beta}^{(b), \text{OC}}_{mj}\ne 0$.
In the following, we work with the bias-reduced two-step estimate  $\pmb{\tilde{\beta}}^{(b), \text{OC}}_{m}$ (see, e.g., \citealp{Belloni2013}) that sets $\tilde{\beta}^{(b), \text{OC}}_{m,j}=0$ if component $j$ was not lasso selected and replaces all other components by estimates from OLS with only the lasso pre-selected components as regressors.

Hence, a straight-forward, non-negative measure of variable importance for $x_j \in \{x_1,...,x_p\}$ is the absolute average of the estimates across all data sets $(m,b)$ computed as 
\begin{equation} \label{meth:eq:im}
	I_j^\text{OC} = \dfrac{1}{MB}\bigg| \sum\limits_{m=1}^{M}  \sum\limits_{b=1}^{B} \tilde{\beta}_{mj}^{(b),\text{OC}} \bigg|\;.
\end{equation}

\subsection*{Step 3: Initial aMIRL Estimates}
For countervailing high correlations among the considered variables, we use only one third of the variables per bootstrap sample in the second step of the modified random lasso procedure. Thus, for each standardized and time-demeaned imputation-bootstrap-pair sample $(m,b)$ we randomly select $\lfloor p/3\rfloor$ candidate variables with selection probability for component $j$ proportional to its importance measure $I_j^\text{OC}$ from~\eqref{meth:eq:im}.\footnote{\cite{Liu2016} select $p/2$ instead of $p/3$ variables, but state that both variants yield similar results.} We perform lasso-OLS on the time-demeaned bootstrapped and imputed data including only the chosen $\lfloor p/3\rfloor$ candidate variables in the potential set of regressors. For each variable $x_j$ we thereby obtain $\tilde{\beta}_{mj}^{(b), \text{OC},\lfloor p/3\rfloor},$ which equals zero if variable $j$ was either not included in the set of potential regressors or not selected by the lasso-OLS procedure. 
The vector of initial aMIRL estimates $\mathbf{\hat{b}}^{\text{init}, \text{OC}}= \left(\hat{b}_1^{\text{init}, \text{OC}}, ..., \hat{b}_p^{\text{init}, \text{OC}}\right)^\prime$ is then computed by averaging the resulting estimates, i.e.
\begin{equation}\label{eq:b_init}
	\hat{b}_j^{\text{init}, \text{OC}} = \dfrac{1}{MB} \sum\limits_{m=1}^{M}  \sum\limits_{b=1}^{B} \tilde{\beta}_{mj}^{(b), \text{OC},\lfloor p/3\rfloor}.
\end{equation}

\subsection*{Step 4: Stability Selection and aMIRL Estimates}
Stability selection, which produces a variable ranking and ultimately the set of variables to be included in the final model, is similar to the random lasso in the previous step, in that we consider lasso estimates based on a subset of the data. Here, too, $\lfloor p/3\rfloor$ randomly selected candidate variables per imputed-bootstrapped data set are used as potential regressors. Instead of employing the OC-optimal tuning parameter $\lambda^\text{OC}$, however, we perform the lasso step over a grid $\Lambda$. 
We then obtain the  empirical selection probabilities $\hat{\pi}_j^\text{OC}$ as
\begin{equation*}
		\hat{\pi}_j^\text{OC} = \underset{\lambda \in \Lambda}{\max} \;  \dfrac{1}{MB} \sum\limits_{m=1}^{M} \sum\limits_{b=1}^{B} \mathds{1} \left\{\hat{\beta}_{mj}^{(b),\lfloor p/3\rfloor, \lambda} \ne 0\right\},
\end{equation*}
where the OC-dependence stems from the fact that the sets of $\lfloor p/3\rfloor$ potential regressors differ for each optimality criterion and pair $(m,b)$ due to the differing information criteria $I_j^\text{OC}$. These selection probabilities introduce a natural importance ranking of all considered variables so that the stable set can be determined as  $\hat{S}^{\text{stable, OC}} = \{ j: \hat{\pi}_j^\text{OC} \ge \pi^{*, \text{OC}} \}$, where $\pi^{*, \text{OC}}$ denotes a data-driven threshold. For more information on the choice of $\pi^{*, \text{OC}}$ and $\Lambda$, see Section~\ref{Method:Sec:OC} below.

The aMIRL estimates $b_{\text{aMIRL},j}^\text{OC}$ are thus only non-zero for the components in the stable set and computed as
\begin{equation}\label{eq:aMIRL}
	b_{\text{aMIRL},j}^\text{OC} = \hat{b}_j^{\text{init, OC}} \times \mathds{1}\left\{j \in\hat{S}^{\text{stable, OC}} \right\}.
\end{equation}

\subsection{Adaptive Tuning Parameter Choice and Evaluation Criteria} \label{Method:Sec:OC}
In this work, we aim to maximise the explanatory power of our models and therefore optimise their in-sample performance with respect to three optimality criteria, namely the Bayesian Information Criterion (BIC), the Akaike Information Criterion (AIC) and Mallows' C$_p$. Since we can retrieve estimates $\hat{\alpha}_i$ for $\alpha_i$ as $\hat{\alpha}_i = \bar{y}_i - \hat{\beta}_1 \bar{x}_{i1} - ... - \hat{\beta}_K \bar{x}_{iK}$ for $i=1,...,N$, see \cite{Wooldridge2012}, we have that $y_{it} - \hat{\alpha}_i - \mathbf{x}_{it}'\pmb{\hat{\beta}} = \ddot{y}_{it} - \mathbf{\ddot{x}}_{it}'\pmb{\hat{\beta}}$. This leads to minimising

\begin{equation*}
  \begin{gathered}
    BIC = NT\,\log\left(\frac{1}{NT} \sum\limits_{i=1}^{N}\sum\limits_{t=1}^{T} (\ddot{y}_{it} - \mathbf{\ddot{x}}_{it}'\pmb{\hat{\beta}})^2\right) + \log(NT)\,(N+K) \\
    AIC = NT\,\log\left(\frac{1}{NT} \sum\limits_{i=1}^{N}\sum\limits_{t=1}^{T} (\ddot{y}_{it} - \mathbf{\ddot{x}}_{it}'\pmb{\hat{\beta}})^2\right) + 2 \,(N+K) \\
    C_p = \dfrac{1}{\hat{\sigma}^2}\sum\limits_{i=1}^{N}\sum\limits_{t=1}^{T} (\ddot{y}_{it} - \mathbf{\ddot{x}}_{it}'\pmb{\hat{\beta}})^2 - NT + 2\,K
  \end{gathered}
\end{equation*}

with $K$ denoting the numbers of regressor variables included in the model, i.e. the number of non-zero components of $\pmb{\hat{\beta}}$, and $\hat{\sigma}^2$ being the variance estimate of the model with all $p$ regressors, i.e. $$\hat{\sigma}^2 = \dfrac{1}{NT-N-p} \sum\limits_{i=1}^{N}\sum\limits_{t=1}^{T} (\ddot{y}_{it} - \mathbf{\ddot{x}}_{it}'\pmb{\hat{\beta}}_\text{full})^2,$$
where $\pmb{\hat{\beta}}_\text{full}$ denotes the within-estimate for $\pmb{\beta}$ of the full model, see \cite{Mallows2000}.\footnote{The definitions of BIC and AIC differ from those of the actual information criteria since the first summand does not equal $-2\,\log(\hat{L})$, $\hat{L}$ being the estimated likelihood of the model, but under normality they have the same minimiser.}

Following \cite{Zheng1995}, we determine the optimal threshold $\pi^{*, \text{OC}}$ via minimising the BIC, more precisely the average BIC across all imputed data sets $m=1,...,M$ over all levels of $\pi$, i.e. $$\pi^{*, \text{OC}} = \underset{\pi \in \{\hat{\pi}_1,...,\hat{\pi}_p\}}{\argmin} \, \dfrac{1}{M} \sum\limits_{m=1}^{M}BIC(m,\pi),$$ where
\begin{equation*}
    BIC(m,\pi) = NT\,\log\left(\frac{1}{NT} \sum\limits_{i=1}^{N}\sum\limits_{t=1}^{T} \left(\ddot{y}_{m,it} - \mathbf{\ddot{x}}_{m,it}'\mathbf{\hat{b}}^{\text{init}, \text{OC}, \pi}\right)^2\right) + \log(NT)\,(N+K)\\
\end{equation*}
with $\mathbf{\hat{b}}^{\text{init}, \text{OC}, \pi} =  \mathbf{\hat{b}}^{\text{init}, \text{OC}}\times \mathds{1}\left\{j : \hat{\pi}_j\ge \pi \right\}$.

Regarding the set of initial tuning parameters $\Lambda=\left\{(\lambda_k)_{k=1}^{K}\right\}$, \cite{Friedman2010} suggest on the log scale decreasing values from $\lambda_\text{max}$ to $\lambda_\text{min}=\delta \lambda_\text{max}$, where $\lambda_\text{max}$ denotes the smallest $\lambda$ for which the estimated coefficient vector $\pmb{\hat{\beta}}$ equals zero. Hence, the sequence is computed as
$\left(\lambda_k\right)_{k=1}^{K}= \exp{s_k},$
where $\left(s_k\right)_{k=1}^{K}$ is an equally spaced sequence from $\log\lambda_\text{max}$ to $\log\lambda_\text{min}$ of length $K$. They propose $K=100$ and $\delta=0.001$, which are also used in this work. We determine 
$\lambda_\text{max} := \underset{m \in M, \,b\in B}{\max}\; \lambda_0^{m, b},$
 where $\lambda_0^{m,b}$ is the smallest $\lambda$ for which the fixed effects lasso estimator of the $b$th bootstrap sample drawn from the $m$th imputed data set equals zero.

 For the evaluation of our results in Section~\ref{Sec:Results}, we compute the significance of the obtained aMIRL estimates via nonparametric, bias-corrected and accelerated bootstrap confidence intervals that do not hinge on any potentially misspecified parametric assumption and are robust to the quality of the asymptotic approximation in finite samples \citep{Efron2020}. In contrast to conventional bootstrap confidence intervals, they allow for (i) nonnormality of the $b_\text{MIRL}$ estimate, (ii) a bias in the aMIRL estimate and (iii) potentially nonconstant standard errors of $b_\text{aMIRL}$. Further details are presented in \cite{Efron2020}. We use jackknife resampling as implemented in the function \texttt{bcajack} in the R package \texttt{bcaboot}. The effect of variable $j$ is significant at the approximate level $\alpha$, if the respective bootstrap confidence interval does not contain zero.

As goodness-of-fit measures, we use the ordinary (adjusted) $R^2$ as well as the (adjusted) within-$R^2$ which is defined as the ordinary (adjusted) $R^2$ of the time-demeaned model. The latter can be interpreted as the amount of time variation in the target variable $y_{it}$ that is explained by the time variation in the predictors ${\mathbf{x}}_{it}$.


\section{Results}\label{Sec:Results}

We apply the aMIRL procedure from Section~\ref{Sec:Method} to the balanced panel data described in Section~\ref{Sec:Data}. For the key financial indicator \textit{OSS} as well as for each of the two social impact target variables \textit{log(NAB)} and \mbox{\textit{--log(ALBG)}}, see Section~\ref{Data:Sec:TarVars} for more details, we run the procedure three times, i.e. once per different optimality criterion BIC, AIC and Mallows' C$_p$. While optimising for BIC yields the smallest models, minimising AIC and C$_p$ results in larger models that are very similar to each other.\footnote{In fact, AIC and C$_p$ have been shown to be equivalent in the case of Gaussian linear regression which may explain the similarity in the results here \citep{Boisbunon2013}.} 
The direction and size of the respective effects per target variable are consistent for all three optimality criteria. Hence, the essence of the models is the same. For simplicity, in this section we report and refer to the estimates for AIC of those variables selected in at least two of the final models. A qualitative overview of the estimated selected effects per factor is given in Table \ref{Results:Tab:Overview} while quantitative results are described in the following two subsections and portrayed in Tables \ref{Results:Tab:OSS}--\ref{Results:Tab:neglogALBG}. Detailed results are provided in Tables \ref{Tab:aMIRL_OSS}--\ref{Tab:aMIRL_neglogALBG} in the appendix.
 
\begin{table}[!htbp]
\centering\footnotesize
\caption{Overview per target variable of the direction and approximate size of the aMIRL coefficients of factors for the AIC optimality citerion where we only focus on variables that are selected by the aMIRL procedure for at least two optimality criteria from the group of AIC, BIC and C$_p$. All respective variables have effects at least of size $\ge 0.01$ in absolute value and are significant at the $1\%$ level based on bootstrap confidence intervals. We roughly distinguish the cases \raisebox{-1pt}{\color{junglegreen}$\mathlarger{\mathlarger{\mathlarger{\bm{\oplus}}}}$}$: b_\text{aMIRL}^{AIC}\ge 0.1$, \raisebox{0pt}{\color{junglegreen}$\bm{\oplus}$}$: 0.1 > b_\text{aMIRL}^{AIC}\ge 0.01$, negative analogously.}
\begin{tabular}{llll}
	\toprule
	\textbf{Factor groups} & \textbf{\textit{OSS}} & \textbf{\textit{log(NAB)}} & \textbf{\textit{--log(ALBG)}} \\ \toprule 
 
	\colorbox{babyblue!70}{\textsc{Costs}} & \raisebox{0pt}{\large\color{junglegreen}$\bm{\oplus}$}\raisebox{0pt}{\large\color{red}$\bm{\ominus\ominus}$}\raisebox{1pt}{\color{red}$\bm{\ominus}$} &  \raisebox{1pt}{\color{junglegreen}$\bm{\oplus}$}\raisebox{1pt}{\color{red}$\bm{\ominus\ominus}$} & \raisebox{0pt}{\large\color{junglegreen}$\bm{\oplus}$}\raisebox{1pt}{\color{junglegreen}$\bm{\oplus}$}\raisebox{0pt}{\large\color{red}$\bm{\ominus}$}\raisebox{1pt}{\color{red}$\bm{\ominus\ominus}$}\\
 
	\colorbox{babyblue!40}{\textsc{Financial Performance}} & \raisebox{0pt}{\large\color{junglegreen}$\bm{\oplus\oplus}$}\raisebox{1pt}{\color{junglegreen}$\bm{\oplus}$} & \raisebox{1pt}{\color{junglegreen}$\bm{\oplus}$}& \raisebox{1pt}{\color{junglegreen}$\bm{\oplus}$}\raisebox{1pt}{\color{red}$\bm{\ominus}$} \\
 
	\colorbox{babyblue!20}{\textsc{Financing Structure}} & & \raisebox{1pt}{\color{red}$\bm{\ominus}$} & \raisebox{0pt}{\color{red}$\bm{\ominus}$} \\
 
	\colorbox{gray!40}{\textsc{Operational Aspects}} & \raisebox{1pt}{\color{junglegreen}$\bm{\oplus\oplus}$}\raisebox{0pt}{\large\color{red}$\bm{\ominus}$} & \raisebox{1pt}{\color{junglegreen}$\bm{\oplus\oplus}$}\raisebox{1pt}{\color{junglegreen}$\bm{\oplus\oplus}$}\raisebox{1pt}{\color{junglegreen}$\bm{\oplus}$}\raisebox{0pt}{\large\color{red}$\bm{\ominus}$}\raisebox{1pt}{\color{red}$\bm{\ominus\ominus}$} & \raisebox{1pt}{\color{junglegreen}$\bm{\oplus\oplus}$}\raisebox{1pt}{\color{junglegreen}$\bm{\oplus\oplus}$}\raisebox{1pt}{\color{junglegreen}$\bm{\oplus\oplus}$}\raisebox{1pt}{\color{red}$\bm{\ominus\ominus}$}\raisebox{1pt}{\color{red}$\bm{\ominus\ominus}$}\raisebox{1pt}{\color{red}$\bm{\ominus\ominus}$}\\
 
	\colorbox{orange!20}{\textsc{Organisational Strategy}} & & \raisebox{1pt}{\color{junglegreen}$\bm{\oplus\oplus}$}\raisebox{1pt}{\color{junglegreen}$\bm{\oplus}$}\raisebox{1pt}{\color{red}$\bm{\ominus}$} & \\
 
	\colorbox{orange!40}{\textsc{Personnel Structure}} & \raisebox{1pt}{\color{red}$\bm{\ominus}$} & \raisebox{0pt}{\large\color{junglegreen}$\bm{\oplus\oplus}$}\raisebox{1pt}{\color{junglegreen}$\bm{\oplus}$}\raisebox{0pt}{\large\color{red}$\bm{\ominus}$}\raisebox{1pt}{\color{red}$\bm{\ominus\ominus}$} & \raisebox{0pt}{\large\color{junglegreen}$\bm{\oplus\oplus}$}\raisebox{0pt}{\large\color{junglegreen}$\bm{\oplus}$}\raisebox{1pt}{\color{junglegreen}$\bm{\oplus}$}\raisebox{0pt}{\large\color{red}$\bm{\ominus}$}\raisebox{1pt}{\color{red}$\bm{\ominus}$}\\
 
    \colorbox{orange!60}{\textsc{Services}} &  & \raisebox{1pt}{\color{junglegreen}$\bm{\oplus}$} & \\\bottomrule
\end{tabular}%
\label{Results:Tab:Overview}
\end{table}

We find that financial success is mostly driven by financial aspects of the MFI while social success depends largely on staff characteristics and operational features. By definition, the operational self-sufficiency of MFIs can be promoted by increasing profitability and reducing costs. The same applies, to a lesser extent, to both dimensions of MFI outreach. The personnel structure plays an important role in all three dimensions of MFI success. While a larger management has a significant negative impact on all three measures of success, changes in the size, efficiency and salary structure of the staff are crucial determinants of greater breadth and depth of outreach. 
Operational aspects are important for social success, but less so for financial success. In general, financial and social success of MFIs are not in opposition to each other which is in line with the findings of \cite{Quayes2015}. The relationship between the two measures of social success is more ambiguous. While issuing smaller loans can contribute to increased breadth of outreach, reaching more borrowers usually leads to targeting wealthier customers, a phenomenon referred to in the literature as \textit{mission drift} \citep{Armendariz_2011}.

\subsection{Key Financial Success Factors} \label{Results:Sec:FinPerf}
By analysing \textit{OSS}, we find that the key characteristics of operational sustainability correspond to financial measures while operational and personnel aspects are of secondary importance, see Table \ref{Results:Tab:OSS} for an overview of the results and Table \ref{Tab:aMIRL_OSS} in the appendix for detailed results of the aMIRL models for all three optimality criteria. Our models yield an $R^2$ of 0.88, which is 0.54 (2.6 times) higher than that of the best model of \cite{Quayes2015}, who also studies \textit{OSS} with a fixed-effects model on a balanced panel of similar size and obtains an $R^2$ of 0.34. Other empirical studies also obtain models with smaller $R^2$ values than ours, see \cite{Kinde2012} (0.58), \cite{2012_Bogan} (0.37), \cite{Strom2014} (0.25) and \cite{Bassem2012} (0.12), amongst others. The variables contained in their models are also included in our potential set of regressors, except for age (which is accounted for indirectly in our setting by the fact that we only consider MFIs that have reported and therefore survived for at least 6 years), type of organisation and country-specific controls (both of which are accounted for by a fixed effect, as they are (likely) constant over time).

\begin{table}[htb]
\caption{Empirical selection probabilities and the aMIRL coefficients of factors for the AIC optimality citerion where we only focus on variables that are selected by the aMIRL procedure for at least two optimality criteria from the group of AIC, BIC and C$_p$ with target variable \textbf{OSS}. All effects are significant at the $1\%$ level based on bootstrap confidence intervals. All reported effects have the same direction in the case of using quantile regression in the post-imputation and post-selection step for the quantile levels $\tau = 0.25,\,0.5,\,0.75$ (see Section \ref{Results:Sec:RobChecks} for details). The rare instances where the direction of at least one of the effects deviates from the aMIRL estimate are marked with coloured bullets $^{\bullet\bullet\bullet}$ with a bullet per quantile level $\tau = 0.25,\,0.5,\,0.75$ where \textcolor{junglegreen}{$^{\bullet}$} indicates a positive, and \textcolor{red}{$^{\bullet}$} marks a negative effect in the respective $\tau$.}
\centering\footnotesize
\begin{tabular}[t]{lcclcc}
\toprule
\textbf{Variable} & $\hat{\pi}_\text{aMIRL}^{AIC}$ & $b_\text{aMIRL}^{AIC}$ & \textbf{Variable} & $\hat{\pi}_\text{aMIRL}^{AIC}$ & $b_\text{aMIRL}^{AIC}$\\
\midrule
\colorbox{babyblue!70}{\textbf{Operating Expense $/$ Assets}} & $\mathbf{0.866}$ & \textcolor{red}{$\mathbf{-0.229}$} & \colorbox{babyblue!45}{Return on Assets} & $0.913$ & $ 0.070$\\
\colorbox{babyblue!70}{Financial Expense $/$ Assets} & $0.982$ & \textcolor{red}{$-0.183$} & \colorbox{gray!40}{Av. Loan Size: Urban} & $0.907$ & \textcolor{red}{$-0.106$\textcolor{red}{$^{\bullet}$}\textcolor{junglegreen}{$^{\bullet}$}\textcolor{junglegreen}{$^{\bullet}$}}\\
\colorbox{babyblue!70}{Operating Expense $/$ GLP} & $0.826$ & $ 0.111$ & \colorbox{gray!40}{Tax Expense $/$ Assets} & $0.872$ & $ 0.059$\\
\colorbox{babyblue!70}{Cost per Borrower} & $0.780$ & \textcolor{red}{$-0.044$} & \colorbox{gray!40}{Write-Off Ratio} & $0.776$ & $ 0.053$\textcolor{red}{$^{\bullet}$}\textcolor{junglegreen}{$^{\bullet}$}\textcolor{junglegreen}{$^{\bullet}$}\\
\colorbox{babyblue!45}{\textbf{Profit Margin}} & $\mathbf{1.000}$ & $\mathbf{ 0.484}$ & \colorbox{orange!45}{\% Staff: Managers} & $0.863$ & \textcolor{red}{$-0.075$\textcolor{red}{$^{\bullet}$}\textcolor{red}{$^{\bullet}$}\textcolor{junglegreen}{$^{\bullet}$}}\\
\colorbox{babyblue!45}{\textbf{Financial Revenue $/$ Assets}} & $\mathbf{0.972}$ & $\mathbf{ 0.263}$ &  &  & \\
\bottomrule
\end{tabular}
\label{Results:Tab:OSS}
\end{table}

Given the definition of \textit{OSS}, it is no surprise that financial performance characteristics and costs are the main determinants of an MFI's operational self-sufficiency as reflected by the five largest estimated effects of the variables \textit{Profit Margin}, \textit{Financial Revenue $/$ Assets}, \textit{Operating Expense $/$ Assets}, \textit{Financial Expense $/$ Assets} and \textit{Operating Expense $/$ GLP}. Similarly, higher relative tax expenses, an indicator of increased taxable capital, are positively associated with operational sustainability. More surprising is the negative role of the percentage of managers in the MFI personnel which is mainly driven by less successful MFIs since the effect is positive for the median and the 75\% quantile. Operationally less sustainable MFIs\footnote{Reminder: The 25\% quantile of \textit{OSS} in our data set is 1.032, see Table \ref{Appendix:Tab:DescrStatsBalanced}, where 1 is the critical point for operational sustainability. Hence, the effects of the 25\% quantile describe the characteristics of MFIs on the verge of being operationally sustainable.} are most likely run by a less effective management, so increasing the size of management in that case has a negative impact. The opposite holds true for financially successful MFIs. We further find that smaller loan sizes for urban borrowers (and hence greater depth of outreach) can improve financial performance. However, this effect is not confirmed by the respective 50\% and 75\%-quantile estimates.

In summary, the operational sustainability of an MFI can be increased by focusing on higher profitability and lower costs and ensuring a staff structure with an effective management of moderate size.

These findings complement and significantly extend the results of previous analyses. We find that most of the variables selected in our final models are not included in the models used in the literature \citep{Bassem2012, 2012_Bogan, Kinde2012, Strom2014, Quayes2015}, highlighting the importance of data-driven variable selection. While \cite{Quayes2015} also recognises a negative influence of cost and loan size on \textit{OSS}, many of previously found significant relationships are not confirmed by our results, which may be caused by omitted variables in these studies. For example, \cite{Kinde2012} finds a significant positive correlation with outreach and a negative one with the proportion of donated capital, \cite{2012_Bogan} analyses the role of capital structure and \cite{Strom2014} that of female leadership, the significance of each of which is not supported by our models.

\subsection{Key Social Success Factors} \label{Results:Sec:SocPerf} 
Changes in the personnel structure of an MFI are \textit{the} most relevant driver of both dimensions of social success, the breadth of outreach (reaching as many borrowers as possible) and the depth of outreach (reaching particularly poor borrowers). Corresponding variables are selected with probability greater than 0.98 and have the largest marginal post-lasso effects, see Tables \ref{Results:Tab:logNAB} and \ref{Results:Tab:neglogALBG} for an overview and Tables \ref{Tab:aMIRL_logNAB} and \ref{Tab:aMIRL_neglogALBG} in the appendix for detailed results of both aMIRL analyses. More specifically, a larger borrower-staff ratio and greater number of employees combined with flat hierarchies are crucial for increased social performance. Again, costs are overall negatively associated with social success, while profitability plays a minor but generally positive role. An increase in liabilities in the capital structure can lead to higher social success, which is reflected by the positive effect of \textit{Fin. Exp. Fund. Liab. $/$ Assets} in both cases and the negative effect of \textit{Capital $/$ Assets} for \textit{log(NAB)}.

Regarding the relationship between the two measures of social success, we find that increased depth of outreach can lead to serving a greater number of borrowers. However, reaching more borrowers typically results in serving wealthier borrowers. This is shown by the combined negative effect of \textit{Av. Loan Bal. $/$ GNI p. c.} and \textit{Av. Outst. Bal. $/$ GNI p. c.} on \textit{log(NAB)} and the negative impact of \textit{\# Active Borrowers} on \textit{--log(ALBG)}. MFIs should therefore carefully choose their primary social objective and tailor their activities accordingly. The two types of social success have some additional specific and unique characteristics, which are discussed in the following two subsections.

To our knowledge, the prominent role of staff structure in the social success of MFIs is new to the literature. We find that our aMIRL results for MFI outreach yield a significantly better fit than related models from other authors that also utilise the MIX Market data set. \cite{2012_Bogan} analyses \textit{log(NAB)} using a linear model with an $R^2$ of 0.72, which is 0.26 smaller than that of our aMIRL models, and \cite{Quayes2012} models \textit{--log(ALBG)} with a linear model that produces an $R^2$ of 0.65, which is 0.32 lower than that of our models. \cite{Quayes2022} use \textit{NAB} and \textit{ALBG} as target variables and reach $R^2$ values of 0.73 and 0.27, which are again much smaller than our $R^2$ values for the logarithm. Other popular studies of MFI outreach do not report their $R^2$ values, making it difficult to compare the goodness of fit of their results with ours, see \cite{2007_Hartarska}, \cite{Hermes2011} and \cite{Awaworyi2020}. Moreover, many of these models contain statistically insignificant regressors, which, again, underlines the need for data-driven model selection.

\subsubsection*{Breadth of Outreach}
\label{Results:Sec:SocPerfBreadth}

\begin{table}[htb]
\caption{Empirical selection probabilities and the aMIRL coefficients of factors for the AIC optimality citerion where we only focus on variables that are selected by the aMIRL procedure for at least two optimality criteria from the group of AIC, BIC and C$_p$ with target variable \textbf{log(NAB)}. All reported effects have the same direction in the case of using quantile regression in the post-imputation and post-selection step for the quantile levels $\tau = 0.25,\,0.5,\,0.75$ (see Section \ref{Results:Sec:RobChecks} for details).}
\centering\footnotesize
\begin{tabular}[t]{p{3.75cm}ccp{3.75cm}cc}
\toprule
\textbf{Variable} & $\hat{\pi}_\text{aMIRL}^{AIC}$ & $b_\text{aMIRL}^{AIC}$ & \textbf{Variable} & $\hat{\pi}_\text{aMIRL}^{AIC}$ & $b_\text{aMIRL}^{AIC}$\\
\midrule
\colorbox{babyblue!70}{Cost per Borrower} & $0.630$ & \textcolor{red}{$-0.024$} & \colorbox{gray!40}{\% GLP: Solidarity Group} & $0.528$ & $ 0.013$\\
\colorbox{babyblue!70}{Financial Expense $/$ Assets} & $0.682$ & \textcolor{red}{$-0.015$} & \colorbox{orange!20}{Incentives: Portf. Qual.} & $0.483$ & $ 0.016$\\
\colorbox{babyblue!70}{Fin. Exp. Fund. Liab. $/$ Assets} & $0.627$ & $ 0.011$ & \colorbox{orange!20}{Target Market: Youth} & $0.485$ & $ 0.014$\\
\colorbox{babyblue!45}{Return on Assets} & $0.742$ & $ 0.024$ & \colorbox{orange!20}{Goals: Health Infrastructure} & $0.524$ & $ 0.012$\\
\colorbox{babyblue!20}{Capital $/$ Assets} & $0.755$ & \textcolor{red}{$-0.053$} & \colorbox{orange!20}{Goals: Econ. Improvement} & $0.508$ & \textcolor{red}{$-0.011$}\\
\colorbox{gray!40}{\textbf{Av. Loan Bal. $/$ GNI p. c.}} & $\mathbf{0.992}$ & \textcolor{red}{$\mathbf{-0.201}$} & \colorbox{orange!45}{\textbf{\# Personnel}} & $\mathbf{0.998}$ & $\mathbf{ 0.270}$\\
\colorbox{gray!40}{Av. Outst. Bal. $/$ GNI p. c.} & $0.917$ & $ 0.097$ & \colorbox{orange!45}{\textbf{Borrowers $/$ Staff Member}} & $\mathbf{0.997}$ & $\mathbf{ 0.229}$\\
\colorbox{gray!40}{\% GLP: Microenterprise} & $0.907$ & \textcolor{red}{$-0.043$} & \colorbox{orange!45}{\% Staff: Board Members} & $0.992$ & \textcolor{red}{$-0.179$}\\
\colorbox{gray!40}{Deposit Accounts $/$ Staff} & $0.661$ & $ 0.032$ & \colorbox{orange!45}{Av. Salary $/$ GNI p. c.} & $0.757$ & $ 0.032$\\
\colorbox{gray!40}{\% Borrowers: Urban} & $0.704$ & \textcolor{red}{$-0.026$} & \colorbox{orange!45}{Personnel Expense $/$ GLP} & $0.710$ & \textcolor{red}{$-0.028$}\\
\colorbox{gray!40}{\% GLP: Enterprise Finance} & $0.823$ & $ 0.023$ & \colorbox{orange!45}{\% Staff: Managers} & $0.505$ & \textcolor{red}{$-0.011$}\\
\colorbox{gray!40}{Loan Impairm. Prov. $/$ Assets} & $0.609$ & $ 0.015$ & \colorbox{orange!70}{Offers Other Fin. Services} & $0.523$ & $ 0.016$\\
\bottomrule
\end{tabular}
\label{Results:Tab:logNAB}
\end{table}

In contrast to the services provided by traditional banks, the success of microfinance loans depends heavily on the support that borrowers receive from the MFI, particularly from the staff \citep{Yunus2009}. Generally, we find that a larger number of borrowers is reached first and foremost through the expansion of microfinance activities as such.
The key components for this are a greater number of employees and their efficient deployment. It is not beneficial to focus on a larger board. The emphasis is to be on additional loan officers who run the microfinance business by directly servicing the borrowers. Each staff member should further be trained to serve more borrowers (\textit{Borrower $/$ Staff}) and be given the right incentives and goals (\textit{Incentives: Portfolio Quality}, \textit{Goals: Health Infrastructure}). More borrowers are also attained by targeting new borrower groups, especially young people, and through offering additional services (\textit{Target Market: Youth}, \textit{Deposit Accounts $/$ Staff}, \textit{Offers Other Fin. Services}). The adjustment of the staff structure should be accompanied by changes in the gross loan portfolio (GLP). It is either to be redistributed or to be increased. The first option can be achieved by reducing the average loan size and by lending to groups instead of individuals. Consequently, the depth of outreach can not only further the financial success of an MFI, but also contribute to its breadth of outreach. Additional GLP, on the other hand, may be financed through liabilities which increases the relative financial expense on funding liabilities. They offer a convenient way to quickly acquire new resources. Here, too, costs are negatively related to the success of the MFI. Their effects can be attributed to economies of scale. Profitability, on the other hand, promotes breath of outreach.

\subsubsection*{Depth of Outreach} \label{Results:Sec:SocPerfDepth}

\begin{table}[htb]
\caption{Empirical selection probabilities and the aMIRL coefficients of factors for the AIC optimality citerion where we only focus on variables that are selected by the aMIRL procedure for at least two optimality criteria from the group of AIC, BIC and C$_p$ with target variable \textbf{--log(ALBG)}. All effects are significant at the $1\%$ level based on bootstrap confidence intervals. Grey font symbolises that the respective variable is not included in the final AIC model, but in both the BIC and the C$_\text{p}$ model. All reported effects have the same direction in the case of using quantile regression in the post-imputation and post-selection step for the quantile levels $\tau = 0.25,\,0.5,\,0.75$ (see Section \ref{Results:Sec:RobChecks} for details). The rare instances where the direction of at least one of the effects deviates from the aMIRL estimate are marked with coloured bullets $^{\bullet\bullet\bullet}$ with a bullet per quantile level $\tau = 0.25,\,0.5,\,0.75$ where \textcolor{junglegreen}{$^{\bullet}$} indicates a positive, and \textcolor{red}{$^{\bullet}$} marks a negative effect in the respective $\tau$.}
\centering\footnotesize
\begin{tabular}[t]{p{4cm}ccp{3.6cm}cc}
\toprule
\textbf{Variable} & $\hat{\pi}_\text{aMIRL}^{AIC}$ & $b_\text{aMIRL}^{AIC}$ & \textbf{Variable} & $\hat{\pi}_\text{aMIRL}^{AIC}$ & $b_\text{aMIRL}^{AIC}$\\
\midrule
\colorbox{babyblue!70}{\textbf{Financial Expense $/$ Assets}} & $\mathbf{0.987}$ & \textcolor{red}{$\mathbf{-0.189}$} & \colorbox{gray!40}{Portfolio at Risk: > 90 Days} & $0.849$ & $ 0.029$\\
\colorbox{babyblue!70}{Fin. Exp. Fund. Liab. $/$ Assets} & $0.982$ & $ 0.180$ & \colorbox{gray!40}{Av. Loan Size: Microenterpr.} & $0.575$ & \textcolor{red}{$-0.029$}\\
\colorbox{babyblue!70}{Cost per Borrower} & $0.830$ & \textcolor{red}{$-0.069$} & \colorbox{gray!40}{\% GLP: Individual} & $0.717$ & $ 0.028$\\
\colorbox{babyblue!70}{Cost per Loan} & $0.680$ & \textcolor{red}{$-0.045$} & \colorbox{gray!40}{Av. Loan Size: Urban} & $0.544$ & \textcolor{red}{$-0.027$}\\
\colorbox{babyblue!70}{Operating Expense $/$ GLP} & $0.640$ & $ 0.037$ & \colorbox{gray!40}{\% Borrowers: Male} & \textcolor{gray!70}{$0.510$} & \textcolor{red}{\textcolor{gray!70}{$-0.022$}}\\
\colorbox{babyblue!45}{Return on Assets} & $0.736$ & $ 0.034$ & \colorbox{gray!40}{\% GLP: Renegotiated Loans} & $0.590$ & \textcolor{red}{$-0.015$}\\
\colorbox{babyblue!45}{Profit Margin} & $0.682$ & \textcolor{red}{$-0.031$} & \colorbox{gray!40}{Borrower Retention Rate} & \textcolor{gray!70}{$0.533$} & \textcolor{gray!70}{$ 0.013$}\\
\colorbox{babyblue!20}{Assets} & $0.865$ & \textcolor{red}{$-0.071$} & \colorbox{orange!45}{\textbf{Borrowers $/$ Staff Member}} & $\mathbf{0.995}$ & $\mathbf{ 0.247}$\\
\colorbox{gray!40}{\# Active Borrowers} & $0.823$ & \textcolor{red}{$-0.091$} & \colorbox{orange!45}{\textbf{Av. Salary $/$ GNI p. c.}} & $\mathbf{0.982}$ & \textcolor{red}{$\mathbf{-0.194}$}\\
\colorbox{gray!40}{Interest Income on GLP $/$ GLP} & $0.776$ & $ 0.048$ & \colorbox{orange!45}{Personnel Expense $/$ GLP} & $0.868$ & $ 0.130$\\
\colorbox{gray!40}{\% GLP: Solidarity Group} & $0.707$ & $ 0.041$ & \colorbox{orange!45}{Personnel Expense $/$ Assets} & $0.887$ & $ 0.102$\\
\colorbox{gray!40}{\% GLP: Village Banking (SHG)} & $0.685$ & $ 0.039$ & \colorbox{orange!45}{\# Personnel} & $0.836$ & $ 0.058$\\
\colorbox{gray!40}{GLP $/$ Total Assets} & $0.636$ & \textcolor{red}{$-0.033$} & \colorbox{orange!45}{\% Staff: Managers} & $0.582$ & \textcolor{red}{$-0.021$}\\
\bottomrule
\end{tabular}
\label{Results:Tab:neglogALBG}
\end{table}

The impact of staff structure on the depth of outreach is determined by largely the same variables as in the analysis of the breadth of outreach. The ratio of borrowers to employees remains decisive while the size of the workforce is reflected to a lesser extent in the number of staff and to a greater extent in the combination of the positive effect of personnel expenses and the negative effect of an increase in average wages. Additional personnel expenses should hence be spent on new staff, especially new loan officers, and not on management, who are assumed to receive a much higher salary. We do not find that greater depth of outreach is necessarily associated with higher costs. The results paint a mixed picture here. While \textit{Financial Expense $/$ Assets}, \textit{Cost per Borrower} and \textit{Cost per Loan} are shown to have a large negative impact, we find that greater costs of funding liabilities and greater \textit{Operating Expense $/$ GLP} are actually positively associated with depth of outreach.

Moreover, as also noted by \cite{Yunus2009}, small loans are usually charged higher interest rates due to their higher default risk. This is confirmed here by the effects of \textit{Interest Income on GLP $/$ GLP}, \textit{Portfolio at Risk: >90 Days} and \textit{\% GLP: Renegotiated Loans}. 
We further find that MFIs expanding their microfinance activity, here reflected by serving more borrowers, increasing the \textit{GLP $/$ Assets} ratio and larger \textit{Assets} can deviate from their main goal to serve particularly poor borrowers by extending larger loan sizes. This phenomenon is called \textit{mission drift} and our findings align with previous results of other authors such as \cite{Armendariz_2011}. Focusing on lending to solidarity groups, villages (self-help groups) and individuals (women over men) rather than small and medium enterprises or large corporations can also increase the depth of outreach. 
Further, we find a mixed relationship with profitability.

Both dimensions of an MFI's social success are achieved by similar means, i.e. mainly by adjusting the personnel structure through an increase of the number of staff, by ensuring that employees can serve more borrowers and an efficient, moderately sized management. To a lesser extent, they are driven by aspects of the size and the distribution of the loan portfolio. While increasing the GLP and targeting new borrower groups furthers the breadth of outreach, depth of outreach is enhanced by focusing on certain groups of borrowers (solidarity groups, self-help groups and women). Lastly, although greater depth of outreach may help an MFI to reach more borrowers, MFIs with a greater breadth of outreach tend to neglect their mission to serve the poorest. Simultaneously increasing both the breadth and depth of outreach can therefore be empirically conflicting goals.

\begin{figure}[!htbp]
    \centering 
    \includegraphics[width=\textwidth]{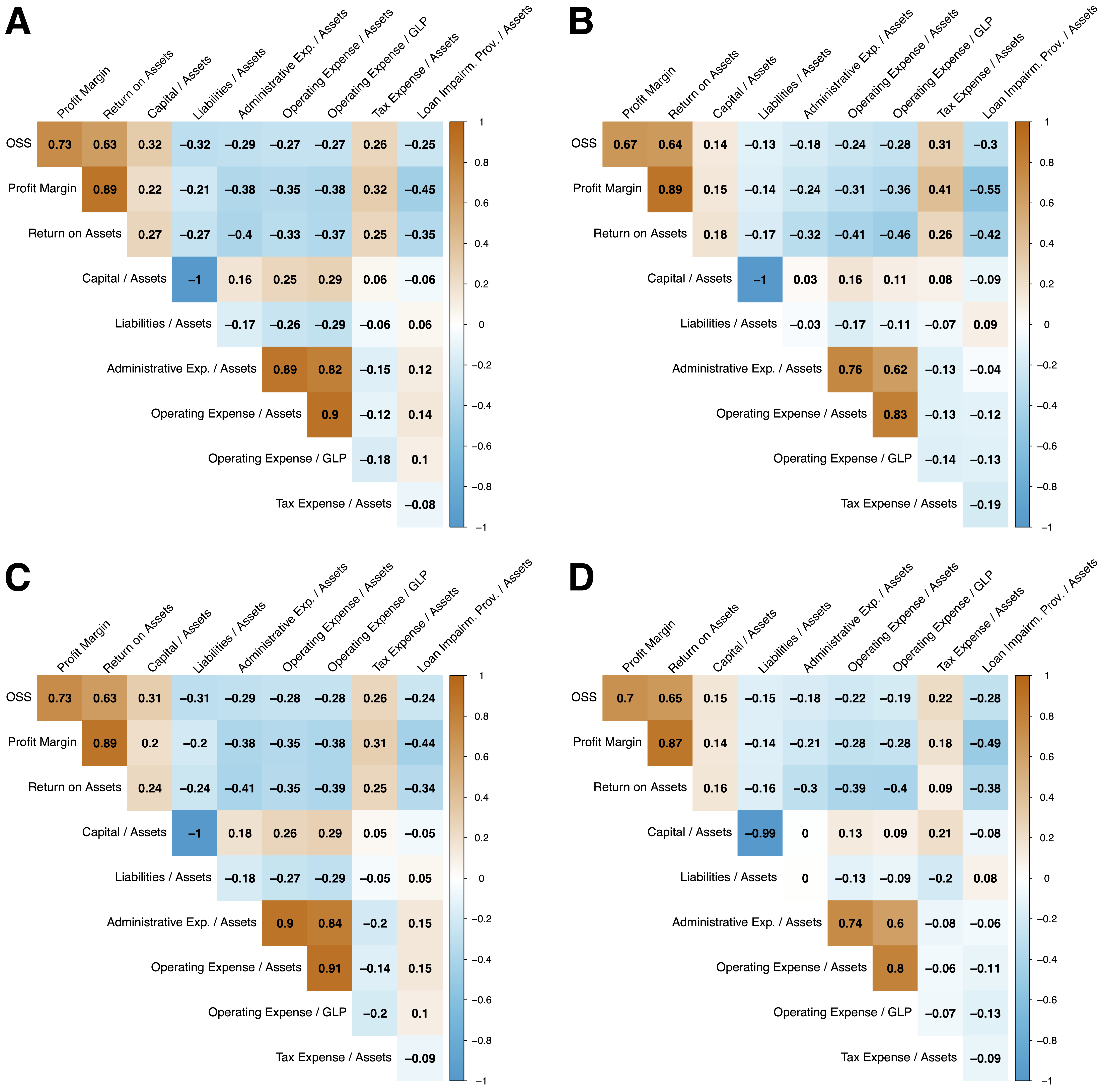}
	\caption[Correlation (standard deviation) for variables most highly correlated with \textit{OSS}]{Correlation (standard deviation) for variables most highly correlated with \textit{OSS}. Pairwise-complete correlation of the cleaned (unimputed) data (top) and average correlation of imputed data sets (bottom). The respective values are computed on the regular (left) and the standardised, time-demeaned data (right). All standard deviations of the correlations of the imputed data sets are between 0 and 0.02.}
	\label{Results:Fig:OSS_corr_sd}
\end{figure}

\subsection{Robustness} \label{Results:Sec:RobChecks}
We compare our aMIRL results with those derived from applying conventional lasso-OLS, with and without fixed effects, as well as the original MIRL algorithm, which does not take into account the panel structure in the data. We also examine the quality of the imputed data, contrast aMIRL estimates for similar outcome variables and compute additional post-selection quantile estimates. We find that the method comparisons confirm our findings while the lasso method with fixed effects is unable to perform robust variable selection in our setting and that the remaining considered methods do not adequately account for the heterogeneity present in the data. The means of two comparable outcome variables are driven by similar factors and quantile estimates confirm that the driving effects are similar across the distribution of the target variables.

\subsubsection*{Comparison to Results from Different Model Selection and Imputation Techniques}
We compare the results obtained from our aMIRL approach with those derived from conventional lasso-OLS estimation. The latter is not robust to multicollinearity and high-dimensionality and does not specifically address the missingness problem. We apply lasso-OLS to the mean-imputed and within-transformed data. Thereby, the approach accounts for unobserved fixed effects in the variable selection and estimation, but not in the imputation step. We find that lasso-OLS yields much larger models than the aMIRL method while producing comparable or up to 13\% (16\%) smaller within (adjusted) $R^2$, see the left columns in Tables \ref{Tab:lasso_OSS}--\ref{Tab:lasso_neglogALBG} in the online supplementary material. The majority of variables selected by the aMIRL procedure are also included in the corresponding lasso model with comparable effect sizes, which affirms our initial findings. However, the majority of the additional variables included in the lasso models are insignificant. Furthermore, the model sizes differ significantly for the different optimality criteria, rendering the results highly sensitive to the choice of the OC.

We also estimate the original MIRL procedure, which uses a common intercept for all observations instead of MFI-specific fixed effects. It is robust to high-dimensionality and multicollinearity, but neither the imputation step, where we use classification and regression trees, nor the estimation and variable selection steps specifically account for the heterogeneity introduced by the panel structure of the data. The results are presented in the middle columns of Tables \ref{Tab:MIRL_OSS}--\ref{Tab:MIRL_neglogALBG} in the appendix. We find that they confirm the main findings from our aMIRL results. However, while the MIRL models contain slightly more time-varying regressors, their overall $R^2$ is much lower (up to 25\%). 

\subsubsection*{Comparison to Results from the Unbalanced Panel}
As the construction of the balanced panel has considerably decreased the sample size available from the unbalanced raw data, we benchmark our empirical results to results from a corresponding unbalanced panel as detailed in Section~\ref{Sec:Data}. In this case, the proposed aMIRL method cannot be used. This is because the within-transformation would lead to the elimination of MFIs that only have observations in one year. Further, for MFIs with very few observations, the time-demeaning step can be problematic.

The standard MIRL method, however, does not hinge on fixed effects imputation and can therefore be applied to deal with the substantial missingness also in the unbalanced panel data (see Section~\ref{Sec:Data} for details). The corresponding MIRL estimates are given in the right column of Tables \ref{Tab:MIRL_OSS}--\ref{Tab:MIRL_neglogALBG}. The overall $R^2$ for the target variables \textit{OSS} and \textit{--log(ALBG)} are approximately 0.07 and 0.5, respectively, and therefore much smaller than those of the balanced panel. This suggests that there is an even larger amount of heterogeneity in the unbalanced panel that is not captured adequately by the MIRL procedure. Moreover, the same linear model with a single intercept for all observations is estimated using lasso-OLS. Consequently, the method does not account for the presence of high-dimensionality, multicollinearity, the panel structure or the missingness of the data. The number of variables included in the final models varies considerably across the target variables, the OC, and the data sets, ranging from six to a high of 123. This is illustrated in the middle and right columns of Tables \ref{Tab:lasso_OSS}--\ref{Tab:lasso_neglogALBG}.

Lastly, since only 58 (177) of all 1278 (3846) observations are fully observed in the balanced (unbalanced) panel data, a comparison with complete case results is infeasible.

Overall, we conclude that there is considerable heterogeneity in the balanced panel, and even more so in the unbalanced case. In combination with the high-dimensionality and multicollinearity inherent to the set of potential regressors as well as the pronounced missingness that cannot be ignored, this renders both the MIRL and the lasso-OLS method incapable of providing robust results. Conversely, the aMIRL results are corroborated by the results of the other models.

\subsubsection*{Imputation Quality} 
As mentioned in Section \ref{Method:Sec:Step1}, the first step of the aMIRL algorithm, i.e. the MICE procedure, is carried out $M=10$ times on the balanced incomplete panel data from Section \ref{Sec:Data}, resulting in ten different imputed data sets. In order to assess the MICE's capacity to maintain linear effects within the data, the pairwise-complete correlations between selected variables in the original data are plotted against the respective average correlations in the imputed data, see Figure~\ref{Results:Fig:OSS_corr_sd}.
We depict those variables that are most highly correlated with \textit{OSS} in the imputed, time-demeaned data set. The plots imply that the imputed data sets contain the same linear structure as the original data as the pairwise-complete and average correlations are very similar. Due to the fact that the standard deviations of the correlations in the imputed data are close to zero, each imputed data set seems to capture the linear dependencies within the data similarly well. The same holds true for time-demeaned data.
Since RE-EM and classification trees are used for imputation, which are by nature nonlinear \citep{Sela2012, DeAth2000}, other effects such as nonlinear dependencies within the data are assumed to be represented in the imputed data sets just as adequately.

We conjecture that the most relevant nonlinearities are captured by the imputation with RE-EM and classification trees.

\subsubsection*{Comparison of Similar Target Variables}
\begin{table}[htb]
\caption{Empirical selection probabilities and the aMIRL coefficients of factors for the AIC optimality citerion where we only focus on variables that are selected by the aMIRL procedure for at least two optimality criteria from the group of AIC, BIC and C$_p$ with target variable \textbf{--ALBG}. All effects are significant at the $1\%$ level based on bootstrap confidence intervals. Grey font symbolises that the respective variable is not included in the final AIC model, but in both the BIC and the C$_\text{p}$ model. The coloured bullets $^{\bullet\bullet\bullet}$ indicate the signs (\textcolor{junglegreen}{$^{\bullet}$} positive, \textcolor{red}{$^{\bullet}$} negative) of the quantile regression estimates, see Section \ref{Results:Sec:RobChecks}, for $\tau = 0.25,\,0.5,\,0.75$ where at least one of them differs from that of the aMIRL estimate.}
\centering\footnotesize
\begin{tabular}[t]{p{3.6cm}ccp{3.6cm}cc}
\toprule
\textbf{Variable} & $\hat{\pi}_\text{aMIRL}^{AIC}$ & $b_\text{aMIRL}^{AIC}$ & \textbf{Variable} & $\hat{\pi}_\text{aMIRL}^{AIC}$ & $b_\text{aMIRL}^{AIC}$\\
\midrule
\colorbox{babyblue!70}{Cost per Loan} & $0.965$ & \textcolor{red}{$-0.210$} & \colorbox{gray!40}{\textbf{Gross Loan Portfolio (GLP)}} & $\mathbf{0.923}$ & $\mathbf{ 0.664}$\\
\colorbox{babyblue!70}{Financial Expense $/$ Assets} & $0.846$ & \textcolor{red}{$-0.085$} & \colorbox{gray!40}{Av. Loan Size: Urban} & $0.789$ & \textcolor{red}{$-0.071$}\\
\colorbox{babyblue!70}{Operating Expense $/$ Assets} & $0.746$ & $ 0.082$ & \colorbox{gray!40}{Av. Loan Size: Microenterpr.} & \textcolor{gray!70}{$0.710$} & \textcolor{red}{\textcolor{gray!70}{$-0.050$}}\\
\colorbox{babyblue!70}{Fin. Exp. Fund. Liab. $/$ Assets} & $0.823$ & $ 0.073$ & \colorbox{orange!45}{Borrowers $/$ Staff Member} & $0.942$ & $ 0.193$\\
\colorbox{babyblue!20}{\textbf{Assets}} & $\mathbf{0.761}$ & \textcolor{red}{$\mathbf{-0.361}$} & \colorbox{orange!45}{Personnel Expense $/$ GLP} & $0.901$ & $ 0.155$\\
\colorbox{babyblue!20}{\textbf{Liabilities}} & $\mathbf{0.763}$ & \textcolor{red}{$\mathbf{-0.342}$\textcolor{red}{$^{\bullet}$}\textcolor{red}{$^{\bullet}$}\textcolor{junglegreen}{$^{\bullet}$}} & \colorbox{orange!45}{\# Personnel} & $0.873$ & $ 0.107$\\
\colorbox{babyblue!20}{Equity} & $0.897$ & \textcolor{red}{$-0.137$\textcolor{red}{$^{\bullet}$}\textcolor{junglegreen}{$^{\bullet}$}\textcolor{junglegreen}{$^{\bullet}$}} & \colorbox{orange!45}{Av. Salary $/$ GNI p. c.} & $0.853$ & \textcolor{red}{$-0.094$}\\
\bottomrule
\end{tabular}
\label{Results:Tab:negALBG}
\end{table}

We compare the results obtained for the negative \textit{Average Loan Balance per GNI per Capita} \textit{(--ALBG)} as the target variable with those for the depth of outreach assuming that the percentage changes in the \textit{ALBG} and the corresponding level changes are determined by similar factors. This assumption is supported by our results as the key findings and takeaways of both analyses are very similar. First, the role of the personnel structure is confirmed by almost the same variables. The relative size differences of the respective estimates vary somewhat, but not to an extent that changes the results significantly. Second, costs play a similar role. Third, the same average loan sizes impacting \textit{--log(ALBG)} are shown to be negatively associated with \textit{--ALBG}. Fourthly, the results for \textit{--ALBG} strongly confirm the existence of mission drift, i.e. the negative effects of an increase in the size of the MFI (measured by \textit{Assets} and here also \textit{Liabilities} and \textit{Equity}) on the depth of outreach. In terms of operational aspects, changes in the distribution of the GLP are not the most important factor here, but rather changes in its size.  The (ambiguous) role of profitability aspects on depth of outreach is not substantiated by the analysis of \textit{--ALBG}.

\subsubsection*{Post-Selection Quantile Estimates}
For each of the three final models per target variable we perform additional post-selection fixed effects quantile regressions. We follow the procedure of \cite{Koenker2004} and estimate the 25\% quantile, the median and the 75\% quantile. Differences in the signs of the estimates compared to the aMIRL estimates are indicated in Tables \ref{Results:Tab:OSS}--\ref{Results:Tab:neglogALBG} and \ref{Tab:aMIRL_OSS}--\ref{Tab:aMIRL_neglogALBG}. In almost all cases, the direction of the marginal quantile effects coincides with that of the mean coefficients. In the rare cases where signs deviate, the median still coincides with the direction of the aMIRL effect with only one exception of one factor in the case of Average Loan Balance per GNI per Capita. This indicates robustness across the distribution of the determined impacts.


\section{Conclusions}\label{Sec:Conc}

We propose the aMIRL method, an adaption of the MIRL procedure introduced by \cite{Liu2016}, that 
combines random lasso with multiple imputation and stability selection. It is tailored to perform robust variable selection for incomplete, high-dimensional longitudinal data with highly correlated variables and unobserved fixed effects. The longitudinal structure in the data is accounted for by using RE-EM trees \citep{Sela2012} in the imputation step and including individual-level fixed effects in the model estimation and variable selection steps.

We utilise this method to determine the success factors of microfinance institutions. To the best of our knowledge, we are the first to do so in a data-driven manner. We do not pre-specify the included regressors, nor do we make the choice of variables depend on their level of missingness. Instead, we apply the aMIRL approach to a transparently constructed balanced panel comprising 1278 observations from 213 MFIs across 136 variables over a six-year period. In accordance with the existing literature, we measure financial success by \textit{Operational Self-Sustainability}. The breadth of outreach (logarithm of the \textit{Number of Active Borrowers}) and the depth of outreach (negative of the logarithm of the \textit{Average Loan Balance per GNI per Capita}) are employed as measures for the different aspects of social success. Our results demonstrate superior explanatory power compared to existing models with the same target variables. For instance, our models for OSS exhibit an overall $R^2$ that is 2.6 times larger (0.88 vs. 0.34) than those of \cite{Quayes2015}, who also estimates a fixed effects model on balanced panel data.

Important practical implications for key decision makers of MFIs can be derived from this work. We show that increased financial sustainability can go hand in hand with greater breadth of outreach while the relationship between financial success and depth of outreach is not as pronounced. Our results further confirm the existence of \textit{mission drift} \citep{Armendariz_2011}. That is, in the process of expanding their lending activities, MFIs tend to deviate from their mission to serve the poorest by extending loans to wealthier borrowers. Optimisation for both dimensions of social success therefore  proves to be empirically problematic. 
Still, both breadth and depth of outreach are furthered by similar factors. The employee structure is found to be the strongest determinant of both dimensions of social success. In particular, a larger number of loan officers trained to serve more borrowers is beneficial as well as appropriate incentives for the loan officers and targeting specific (new) borrower groups. Moreover, in line with literature, we find that greater depth of outreach is associated with higher interest rates and default risk \citep{Yunus2009}. Financial success, on the other hand, is furthered mainly through greater financial performance and efficient management. As far as we know, the role of staff as a key factor in MFI success is new to the literature.

A series of robustness checks was conducted to assess the performance of the method and to evaluate the validity of this study's results. A comparison of the aMIRL results with those of the conventional lasso, with and without fixed effects, as well as MIRL estimates on a large, unbalanced panel comprising 3846 observations of 1026 MFIs, reveals that the results of all of these approaches are significantly inferior to those obtained by the aMIRL method. The corresponding estimates are more unstable, more variables are selected (many of which are insignificant), and the goodness of fit is similar or worse than that obtained by aMIRL. Moreover, we show that the correlation structure within the data is preserved by the imputation and that related outcome variables yield similar results. Lastly, post-selection quantile estimates corroborate our findings for the mean.

Our results show that it is crucial to account for MFI-specific heterogeneity through fixed effects, to use data-driven variable selection rather than pre-specified regressors, and to account for missing data in the analysis instead of excluding incomplete observations. They also empirically demonstrate that, under the given conditions of unobserved individual heterogeneity, pronounced missingness, high-dimensionality and multicollinearity, the aMIRL aproach is more flexible and robust than the methods we compare it against.

Further research could facilitate a deeper understanding of the factors that contribute to MFI success. The incorporation of nonlinear effects and studying simultaneous interactions between different outcome variables may foster interesting results.
In addition, variable selection methods for fixed-effects quantile regressions could provide more comprehensive insights into the distribution of MFI success.  However, extending the aMIRL algorithm for quantiles in panels with fixed effects is not straight-forward. This is due to the inherent non-linearity of the quantile, which does not allow for the pre-elimination of fixed effects. This can lead to an excessive number of variables being removed, resulting in an overestimation of individual effects and a failure to account for the inherent variability in the data. We leave the development of a suitable technique for quantiles to a separate statistical paper.

\begin{acks}[Acknowledgments]
We would like to thank Professor Michael Elliott, Joint Editor, the Associate Editor, and two anonymous referees for their invaluable comments and suggestions, which have significantly enhanced the quality of our article. We are further grateful to the participants of the HKMetrics-Workshop, Mannheim, 2022, the FERN Seminar, Karlsruhe, 2022, the 8th Annual Conference of the International Association for Applied Econometrics (IAAE), London, 2022, the Africa Meeting of the Econometric Society (AFES), Nairobi, 2023, and the 28th International Panel Data Conference (IPDC), Amsterdam, 2023, for valuable feedback.

\textit{Conflict of interest:} None declared.
\end{acks}

\begin{acks}[Funding]
Melanie Schienle gratefully acknowledges funding by the Klaus Tschira Foundation and the Helmholtz Association grant COCAP (KA1-Co-10).
\end{acks}

\begin{acks}[Data and Code]
The data underlying this article are available in the World Bank Data Catalog at \url{https://datacatalog.worldbank.org/dataset/mix-market}. Code for replicating the results can be found at \url{https://github.com/lottarueter/aMIRL}.
\end{acks}


\bibliographystyle{imsart-nameyear} 
\bibliography{library}       



\begin{appendix}
\section*{Detailed Results}\label{appn}
\renewcommand{\arraystretch}{1.15}
\begingroup\fontsize{7}{9}\selectfont

\begin{longtable}[t]{p{3.1cm}wc{0.5cm}wc{1.1cm}wc{0.7cm}wc{0.5cm}wc{1.1cm}wc{0.7cm}wc{0.5cm}wc{1.1cm}wc{0.7cm}}
\caption{Summary of the aMIRL estimation results on the balanced panel data for target variable \textbf{\textit{OSS}}. The table contains all variables that were selected to at least one of the three final models per target variable. sd denotes the empirical standard deviation of the initial estimates the aMIRL estimates are computed from, see equation \eqref{eq:b_init}. All effects are significant at the $1\%$ level based on bootstrap confidence intervals, unless otherwise stated: 
$^{**}$ ($^*$) denotes that the respective effect is significant at the 5\% (10\%) level, $^\circ$ implies that it is not significant at the 10\% level. Grey font symbolises that the respective variable is not included in the final model. The coloured bullets $^{\bullet\bullet\bullet}$ indicate the signs (\textcolor{junglegreen}{$^{\bullet}$} positive, \textcolor{red}{$^{\bullet}$} negative) of the quantile regression estimates for $\tau = 0.25,\,0.5,\,0.75$, where at least one of them differs from that of the aMIRL estimate. 
$R_\text{av.}^2$ ($\bar{R}_\text{av.}^2$) denotes the average (adjusted) $R^2$ across all 10 imputed data sets.}\\
\toprule
\label{Tab:aMIRL_OSS}
 & \multicolumn{3}{c}{\textbf{BIC}} & \multicolumn{3}{c}{\textbf{AIC}} & \multicolumn{3}{c}{\textbf{C}$_\textbf{p}$} \\ \cmidrule[0.5pt](lr{.75em}){2-4} \cmidrule[0.5pt](lr{.75em}){5-7} \cmidrule[0.5pt](lr{.75em}){8-10}
 \textbf{Variable} & ${\hat{\pi}}$ & $b_{\text{aMIRL}}$ & sd$_{b_{\text{aMIRL}}}$ & ${\hat{\pi}}$ & ${b_{\text{aMIRL}}}$ & sd$_{b_{\text{aMIRL}}}$ & ${\hat{\pi}}$ & ${b_{\text{aMIRL}}}$ & sd$_{b_{\text{aMIRL}}}$\\
\midrule
\endfirsthead
\caption{Summary of the aMIRL estimation results on the balanced panel data for target variable \textbf{\textit{OSS}}. \textit{(continued)}}\\
\toprule
  & \multicolumn{3}{c}{\textbf{BIC}} & \multicolumn{3}{c}{\textbf{AIC}} & \multicolumn{3}{c}{\textbf{C}$_\textbf{p}$} \\
 \cmidrule[0.5pt](lr{.75em}){2-4} \cmidrule[0.5pt](lr{.75em}){5-7} \cmidrule[0.5pt](lr{.75em}){8-10}
 \textbf{Variable} & ${\hat{\pi}}$ & $b_{\text{aMIRL}}$ & sd$_{b_{\text{aMIRL}}}$ & ${\hat{\pi}}$ & ${b_{\text{aMIRL}}}$ & sd$_{b_{\text{aMIRL}}}$ & ${\hat{\pi}}$ & ${b_{\text{aMIRL}}}$ & sd$_{b_{\text{aMIRL}}}$\\
\midrule
\endhead

\bottomrule
\multicolumn{10}{r}{(To be continued)}
\endfoot
\bottomrule
\endlastfoot
\multicolumn{10}{l}{\textbf{\textsc{Costs}}}\\
\colorbox{babyblue!70}{Cost per Borrower} & \textcolor{gray!70}{$0.784$} & \textcolor{red}{\textcolor{gray!70}{$-0.016$}} & \textcolor{gray!70}{$0.031$} & $0.780$ & \textcolor{red}{$-0.044$} & $0.061$ & $0.799$ & \textcolor{red}{$-0.041$} & $0.061$\\
\colorbox{babyblue!70}{Fin. Exp. Fund. Liab. $/$ Assets} & \textcolor{gray!70}{$0.188$} & \textcolor{gray!70}{$ 0.002$} & \textcolor{gray!70}{$0.027$} & $0.701$ & $ 0.051$\textcolor{red}{$^{\bullet}$}\textcolor{red}{$^{\bullet}$}\textcolor{red}{$^{\bullet}$} & $0.081$ & \textcolor{gray!70}{$0.622$} & \textcolor{gray!70}{$ 0.043$} & \textcolor{gray!70}{$0.077$}\\
\colorbox{babyblue!70}{Financial Expense $/$ Assets} & $0.998$ & \textcolor{red}{$-0.095$} & $0.071$ & $0.982$ & \textcolor{red}{$-0.183$} & $0.108$ & $0.980$ & \textcolor{red}{$-0.172$} & $0.107$\\
\colorbox{babyblue!70}{Operating Expense $/$ Assets} & \textcolor{gray!70}{$0.696$} & \textcolor{red}{\textcolor{gray!70}{$-0.042$}} & \textcolor{gray!70}{$0.154$} & $0.866$ & \textcolor{red}{${-0.229}$} & $0.232$ & $0.864$ & \textcolor{red}{${-0.211}$} & $0.230$\\
\colorbox{babyblue!70}{Operating Expense $/$ GLP} & \textcolor{gray!70}{$0.891$} & \textcolor{gray!70}{$ 0.053$} & \textcolor{gray!70}{$0.126$} & $0.826$ & $ 0.111$ & $0.175$ & $0.836$ & $ 0.102$ & $0.171$\\
\midrule

\multicolumn{10}{l}{\textbf{\textsc{Financial Performance}}}\\
\colorbox{babyblue!45}{Financial Revenue $/$ Assets} & \textcolor{gray!70}{$0.931$} & \textcolor{gray!70}{$ 0.064$} & \textcolor{gray!70}{$0.161$} & $0.972$ & ${0.263}$ & $0.201$ & $0.970$ & ${0.243}$ & $0.209$\\
\end{longtable}
\endgroup{}

\begingroup\fontsize{7}{9}\selectfont

\begin{longtable}[t]{p{3.1cm}wc{0.5cm}wc{1.1cm}wc{0.7cm}wc{0.5cm}wc{1.1cm}wc{0.7cm}wc{0.5cm}wc{1.1cm}wc{0.7cm}}
\caption{Summary of the aMIRL estimation results on the balanced panel for target variable \textbf{\textit{log(NAB)}}. The table contains all variables that were selected to at least one of the three final models per target variable. sd denotes the empirical standard deviation of the initial estimates the aMIRL estimates are computed from, see equation \eqref{eq:b_init}. All effects are significant at the $1\%$ level based on bootstrap confidence intervals, unless otherwise stated:
$^{**}$ ($^*$) denotes that the respective effect is significant at the 5\% (10\%) level, $^\circ$ implies that it is not significant at the 10\% level. Grey font symbolises that the respective variable is not included in the final model. The coloured bullets $^{\bullet\bullet\bullet}$ indicate the signs (\textcolor{junglegreen}{$^{\bullet}$} positive, \textcolor{red}{$^{\bullet}$} negative) of the quantile regression estimates for $\tau = 0.25,\,0.5,\,0.75$, where at least one of them differs from that of the aMIRL estimate. 
$R_\text{av.}^2$ ($\bar{R}_\text{av.}^2$) denotes the average (adjusted) $R^2$ across all 10 imputed data sets.}\\
\toprule
\label{Tab:aMIRL_logNAB}
 & \multicolumn{3}{c}{\textbf{BIC}} & \multicolumn{3}{c}{\textbf{AIC}} & \multicolumn{3}{c}{\textbf{C}$_\textbf{p}$} \\ \cmidrule[0.5pt](lr{.75em}){2-4} \cmidrule[0.5pt](lr{.75em}){5-7} \cmidrule[0.5pt](lr{.75em}){8-10}
 \textbf{Variable} & ${\hat{\pi}}$ & $b_{\text{aMIRL}}$ & sd$_{b_{\text{aMIRL}}}$ & ${\hat{\pi}}$ & ${b_{\text{aMIRL}}}$ & sd$_{b_{\text{aMIRL}}}$ & ${\hat{\pi}}$ & ${b_{\text{aMIRL}}}$ & sd$_{b_{\text{aMIRL}}}$\\
\midrule
\endfirsthead
\caption[]{Summary of the aMIRL estimation results on the balanced panel data for target variable \textbf{\textit{log(NAB)}}. \textit{(continued)}}\\
\toprule
 & \multicolumn{3}{c}{\textbf{BIC}} & \multicolumn{3}{c}{\textbf{AIC}} & \multicolumn{3}{c}{\textbf{C}$_\textbf{p}$} \\ \cmidrule[0.5pt](lr{.75em}){2-4} \cmidrule[0.5pt](lr{.75em}){5-7} \cmidrule[0.5pt](lr{.75em}){8-10}
 \textbf{Variable} & ${\hat{\pi}}$ & $b_{\text{aMIRL}}$ & sd$_{b_{\text{aMIRL}}}$ & ${\hat{\pi}}$ & ${b_{\text{aMIRL}}}$ & sd$_{b_{\text{aMIRL}}}$ & ${\hat{\pi}}$ & ${b_{\text{aMIRL}}}$ & sd$_{b_{\text{aMIRL}}}$\\
\midrule
\endhead

\bottomrule
\multicolumn{10}{r}{(To be continued)}
\endfoot
\bottomrule
\endlastfoot

\multicolumn{10}{l}{\textbf{\textsc{Costs}}}\\
\colorbox{babyblue!70}{Cost per Borrower} & $0.850$ & \textcolor{red}{$-0.026$} & $0.028$ & $0.630$ & \textcolor{red}{$-0.024$} & $0.027$ & $0.688$ & \textcolor{red}{$-0.028$} & $0.028$\\
\colorbox{babyblue!70}{Fin. Exp. Fund. Liab. $/$ Assets} & \textcolor{gray!70}{$0.242$} & \textcolor{gray!70}{$ 0.001$} & \textcolor{gray!70}{$0.003$} & $0.627$ & $ 0.011$ & $0.024$ & $0.568$ & $ 0.009$ & $0.021$\\
\colorbox{babyblue!70}{Financial Expense $/$ Assets} & \textcolor{gray!70}{$0.056$} & \textcolor{gray!70}{$ 0.000$} & \textcolor{gray!70}{$0.002$} & $0.682$ & \textcolor{red}{$-0.015$} & $0.032$ & $0.620$ & \textcolor{red}{$-0.012$} & $0.028$\\
\midrule

\multicolumn{10}{l}{\textbf{\textsc{Financial Performance}}}\\
\colorbox{babyblue!45}{Return on Assets} & $0.857$ & $ 0.026$ & $0.018$ & $0.742$ & $ 0.024$ & $0.022$ & $0.716$ & $ 0.024$ & $0.021$\\
\colorbox{babyblue!45}{Yield on GLP (Real)} & \textcolor{gray!70}{$0.040$} & \textcolor{red}{\textcolor{gray!70}{$ 0.000^\circ$}} & \textcolor{gray!70}{$0.001$} & $0.530$ & $ 0.005$ & $0.012$ & \textcolor{gray!70}{$0.479$} & \textcolor{gray!70}{$ 0.005$} & \textcolor{gray!70}{$0.012$}\\
\midrule

\multicolumn{10}{l}{\textbf{\textsc{Financing Structure}}}\\
\end{longtable}
\endgroup{}

\renewcommand{\arraystretch}{1.15}

\begingroup\fontsize{7}{9}\selectfont

\begin{longtable}[t]{p{3.1cm}wc{0.5cm}wc{1.1cm}wc{0.7cm}wc{0.5cm}wc{1.1cm}wc{0.7cm}wc{0.5cm}wc{1.1cm}wc{0.7cm}}
\caption{Summary of the aMIRL estimation results on the balanced panel for target variable \textbf{\textit{--log(ALBG)}}. The table contains all variables that were selected to at least one of the three final models per target variable. sd denotes the empirical standard deviation of the initial estimates the aMIRL estimates are computed from, see equation \eqref{eq:b_init}. All effects are significant at the $1\%$ level based on bootstrap confidence intervals, unless otherwise stated: 
$^{**}$ ($^*$) denotes that the respective effect is significant at the 5\% (10\%) level, $^\circ$ implies that it is not significant at the 10\% level. Grey font symbolises that the respective variable is not included in the final model. The coloured bullets $^{\bullet\bullet\bullet}$ indicate the signs (\textcolor{junglegreen}{$^{\bullet}$} positive, \textcolor{red}{$^{\bullet}$} negative) of the quantile regression estimates for $\tau = 0.25,\,0.5,\,0.75$, where at least one of them differs from that of the aMIRL estimate. 
$R_\text{av.}^2$ ($\bar{R}_\text{av.}^2$) denotes the average (adjusted) $R^2$ across all 10 imputed data sets.}\\
\toprule
\label{Tab:aMIRL_neglogALBG}
 & \multicolumn{3}{c}{\textbf{BIC}} & \multicolumn{3}{c}{\textbf{AIC}} & \multicolumn{3}{c}{\textbf{C}$_\textbf{p}$} \\ \cmidrule[0.5pt](lr{.75em}){2-4} \cmidrule[0.5pt](lr{.75em}){5-7} \cmidrule[0.5pt](lr{.75em}){8-10}
 \textbf{Variable} & ${\hat{\pi}}$ & $b_{\text{aMIRL}}$ & sd$_{b_{\text{aMIRL}}}$ & ${\hat{\pi}}$ & ${b_{\text{aMIRL}}}$ & sd$_{b_{\text{aMIRL}}}$ & ${\hat{\pi}}$ & ${b_{\text{aMIRL}}}$ & sd$_{b_{\text{aMIRL}}}$\\
\midrule
\endfirsthead
\caption{Summary of the aMIRL estimation results on the balanced panel data for target variable \textbf{\textit{--log(ALBG)}}. \textit{(continued)}}\\
\toprule
 & \multicolumn{3}{c}{\textbf{BIC}} & \multicolumn{3}{c}{\textbf{AIC}} & \multicolumn{3}{c}{\textbf{C}$_\textbf{p}$} \\ \cmidrule[0.5pt](lr{.75em}){2-4} \cmidrule[0.5pt](lr{.75em}){5-7} \cmidrule[0.5pt](lr{.75em}){8-10}
 \textbf{Variable} & ${\hat{\pi}}$ & $b_{\text{aMIRL}}$ & sd$_{b_{\text{aMIRL}}}$ & ${\hat{\pi}}$ & ${b_{\text{aMIRL}}}$ & sd$_{b_{\text{aMIRL}}}$ & ${\hat{\pi}}$ & ${b_{\text{aMIRL}}}$ & sd$_{b_{\text{aMIRL}}}$\\
\midrule
\endhead

\bottomrule
\multicolumn{10}{r}{(To be continued)}
\endfoot
\bottomrule
\endlastfoot
\multicolumn{10}{l}{\textbf{\textsc{Costs}}}\\
\colorbox{babyblue!70}{Cost per Borrower} & $0.927$ & \textcolor{red}{$-0.057$} & $0.048$ & $0.830$ & \textcolor{red}{$-0.069$} & $0.059$ & $0.859$ & \textcolor{red}{$-0.071$} & $0.056$\\
\colorbox{babyblue!70}{Cost per Loan} & $0.944$ & \textcolor{red}{$-0.060$} & $0.030$ & $0.680$ & \textcolor{red}{$-0.045$} & $0.043$ & $0.748$ & \textcolor{red}{$-0.045$} & $0.041$\\
\colorbox{babyblue!70}{Fin. Exp. Fund. Liab. $/$ Assets} & $0.998$ & $ 0.171$ & $0.060$ & $0.982$ & $ 0.180$ & $0.053$ & $0.991$ & $ 0.184$ & $0.050$\\
\colorbox{babyblue!70}{Financial Expense $/$ Assets} & $0.997$ & \textcolor{red}{${-0.179}$} & $0.065$ & $0.987$ & \textcolor{red}{${-0.189}$} & $0.058$ & $0.992$ & \textcolor{red}{${-0.193}$} & $0.055$\\
\colorbox{babyblue!70}{Operating Expense $/$ GLP} & \textcolor{gray!70}{$0.437$} & \textcolor{gray!70}{$ 0.009$} & \textcolor{gray!70}{$0.025$} & $0.640$ & $ 0.037$ & $0.083$ & $0.647$ & $ 0.034$ & $0.074$\\
\midrule

\multicolumn{10}{l}{\textbf{\textsc{Financial Performance}}}\\
\colorbox{babyblue!45}{Profit Margin} & $0.779$ & \textcolor{red}{$-0.021$} & $0.029$ & $0.682$ & \textcolor{red}{$-0.031$} & $0.039$ & $0.724$ & \textcolor{red}{$-0.031$} & $0.038$\\
\end{longtable}
\endgroup{}

\begin{landscape}

\begingroup\fontsize{7}{9}\selectfont

\begin{longtable}[t]{p{3.3cm}wc{0.5cm}wc{0.8cm}wc{0.5cm}wc{0.5cm}wc{0.8cm}wc{0.5cm}wc{0.5cm}wc{0.8cm}wc{0.5cm}wc{0.5cm}wc{0.8cm}wc{0.5cm}wc{0.5cm}wc{0.8cm}wc{0.5cm}wc{0.5cm}wc{0.8cm}wc{0.5cm}wc{0.5cm}wc{0.8cm}wc{0.5cm}}
\caption{Summary of the MIRL estimation results on the balanced and unbalanced panel data for target variable \textbf{\textit{OSS}}. The table contains all variables that were selected to at least one of the three final models per target variable. sd denotes the empirical standard deviation of the initial estimates the MIRL estimates are computed from, see equation \eqref{eq:b_init}. All effects are significant at the $1\%$ level based on bootstrap confidence intervals, unless otherwise stated: 
$^{**}$ ($^*$) denotes that the respective effect is significant at the 5\% (10\%) level, $^\circ$ implies that it is not significant at the 10\% level. Grey font symbolises that the respective variable is not included in the final model. 
$R_\text{av.}^2$ ($\bar{R}_\text{av.}^2$) denotes the average (adjusted) $R^2$ across all 10 imputed data sets.}\\
\toprule
\label{Tab:MIRL_OSS}
 & \multicolumn{9}{c}{Balanced Panel} & \multicolumn{9}{c}{Unbalanced Panel}\\
 \cmidrule[0.5pt](lr{.75em}){2-10} \cmidrule[0.5pt](lr{.75em}){11-19}

 & \multicolumn{3}{c}{\textbf{BIC}} & \multicolumn{3}{c}{\textbf{AIC}} & \multicolumn{3}{c}{\textbf{C}$_\textbf{p}$} & 
  \multicolumn{3}{c}{\textbf{BIC}} & 
  \multicolumn{3}{c}{\textbf{AIC}} & \multicolumn{3}{c}{\textbf{C}$_\textbf{p}$}
  \\ \cmidrule[0.5pt](lr{.75em}){2-4} \cmidrule[0.5pt](lr{.75em}){5-7} \cmidrule[0.5pt](lr{.75em}){8-10} \cmidrule[0.5pt](lr{.75em}){11-13} \cmidrule[0.5pt](lr{.75em}){14-16} \cmidrule[0.5pt](lr{.75em}){17-19}
 \textbf{Variable} & ${\hat{\pi}}$ & $b_{\text{MIRL}}$ & sd$_{b_{\text{MIRL}}}$ & ${\hat{\pi}}$ & ${b_{\text{MIRL}}}$ & sd$_{b_{\text{MIRL}}}$ & ${\hat{\pi}}$ & ${b_{\text{MIRL}}}$ & sd$_{b_{\text{MIRL}}}$ & ${\hat{\pi}}$ & $b_{\text{MIRL}}$ & sd$_{b_{\text{MIRL}}}$ & ${\hat{\pi}}$ & ${b_{\text{MIRL}}}$ & sd$_{b_{\text{MIRL}}}$ & ${\hat{\pi}}$ & ${b_{\text{MIRL}}}$ & sd$_{b_{\text{MIRL}}}$\\
\midrule
\endfirsthead
\caption{Summary of the MIRL estimation results based on the balanced and unbalanced panel data for target variable \textbf{\textit{OSS}}. \textit{(continued)}}\\
\toprule
 & \multicolumn{9}{c}{Balanced Panel} & \multicolumn{9}{c}{Unbalanced Panel}\\ \cmidrule[0.5pt](lr{.75em}){2-10} \cmidrule[0.5pt](lr{.75em}){11-19}

 & \multicolumn{3}{c}{\textbf{BIC}} & \multicolumn{3}{c}{\textbf{AIC}} & \multicolumn{3}{c}{\textbf{C}$_\textbf{p}$} & 
  \multicolumn{3}{c}{\textbf{BIC}} & 
  \multicolumn{3}{c}{\textbf{AIC}} & \multicolumn{3}{c}{\textbf{C}$_\textbf{p}$}
  \\ \cmidrule[0.5pt](lr{.75em}){2-4} \cmidrule[0.5pt](lr{.75em}){5-7} \cmidrule[0.5pt](lr{.75em}){8-10} \cmidrule[0.5pt](lr{.75em}){11-13} \cmidrule[0.5pt](lr{.75em}){14-16} \cmidrule[0.5pt](lr{.75em}){17-19}
 \textbf{Variable} & ${\hat{\pi}}$ & $b_{\text{MIRL}}$ & sd$_{b_{\text{MIRL}}}$ & ${\hat{\pi}}$ & ${b_{\text{MIRL}}}$ & sd$_{b_{\text{MIRL}}}$ & ${\hat{\pi}}$ & ${b_{\text{MIRL}}}$ & sd$_{b_{\text{MIRL}}}$ & ${\hat{\pi}}$ & $b_{\text{MIRL}}$ & sd$_{b_{\text{MIRL}}}$ & ${\hat{\pi}}$ & ${b_{\text{MIRL}}}$ & sd$_{b_{\text{MIRL}}}$ & ${\hat{\pi}}$ & ${b_{\text{MIRL}}}$ & sd$_{b_{\text{MIRL}}}$\\
\midrule
\endhead

\bottomrule
\multicolumn{19}{r}{(To be continued)}
\endfoot
\bottomrule
\endlastfoot
\multicolumn{19}{l}{\textbf{\textsc{Costs}}}\\
\colorbox{babyblue!70}{Cost per Borrower} & \textcolor{gray!70}{$0.019$} & \textcolor{gray!70}{$ 0.000^\circ$} & \textcolor{gray!70}{$0.000$} & $0.963$ & $ 0.225$ & $0.172$ & $0.953$ & $ 0.232$ & $0.170$ & \textcolor{gray!70}{$0.051$} & \textcolor{red}{\textcolor{gray!70}{$ 0.000$}} & \textcolor{gray!70}{$0.007$} & \textcolor{gray!70}{$0.384$} & \textcolor{gray!70}{$ 0.016$} & \textcolor{gray!70}{$0.121$} & \textcolor{gray!70}{$0.400$} & \textcolor{gray!70}{$ 0.016$} & \textcolor{gray!70}{$0.126$}\\
\colorbox{babyblue!70}{Cost per Loan} & $0.986$ & \textcolor{red}{$-0.037$} & $0.032$ & $0.997$ & \textcolor{red}{$-0.283$} & $0.171$ & $0.993$ & \textcolor{red}{$-0.290$} & $0.169$ & \textcolor{gray!70}{$0.433$} & \textcolor{red}{\textcolor{gray!70}{$-0.003$}} & \textcolor{gray!70}{$0.015$} & \textcolor{gray!70}{$0.723$} & \textcolor{red}{\textcolor{gray!70}{$-0.029$}} & \textcolor{gray!70}{$0.120$} & \textcolor{gray!70}{$0.705$} & \textcolor{red}{\textcolor{gray!70}{$-0.029$}} & \textcolor{gray!70}{$0.124$}\\
\colorbox{babyblue!70}{Fin. Exp. Fund. Liab. $/$ Assets} & \textcolor{gray!70}{$0.145$} & \textcolor{red}{\textcolor{gray!70}{$ 0.000$}} & \textcolor{gray!70}{$0.006$} & $0.839$ & $ 0.122$ & $0.113$ & $0.810$ & $ 0.118$ & $0.112$ & \textcolor{gray!70}{$0.841$} & \textcolor{red}{\textcolor{gray!70}{$-0.022$}} & \textcolor{gray!70}{$0.057$} & \textcolor{gray!70}{$0.743$} & \textcolor{red}{\textcolor{gray!70}{$-0.039$}} & \textcolor{gray!70}{$0.075$} & \textcolor{gray!70}{$0.732$} & \textcolor{red}{\textcolor{gray!70}{$-0.036$}} & \textcolor{gray!70}{$0.069$}\\
\end{longtable}
\endgroup{}

\begingroup\fontsize{7}{9}\selectfont

\begin{longtable}[t]{p{3.3cm}wc{0.5cm}wc{0.8cm}wc{0.5cm}wc{0.5cm}wc{0.8cm}wc{0.5cm}wc{0.5cm}wc{0.8cm}wc{0.5cm}wc{0.5cm}wc{0.8cm}wc{0.5cm}wc{0.5cm}wc{0.8cm}wc{0.5cm}wc{0.5cm}wc{0.8cm}wc{0.5cm}wc{0.5cm}wc{0.8cm}wc{0.5cm}}
\caption{Summary of the MIRL estimation results on the balanced and unbalanced panel data for target variable \textbf{\textit{log(NAB)}}. The table contains all variables that were selected to at least one of the three final models per target variable. sd denotes the empirical standard deviation of the initial estimates the MIRL estimates are computed from, see equation \eqref{eq:b_init}. All effects are significant at the $1\%$ level based on bootstrap confidence intervals, unless otherwise stated: 
$^{**}$ ($^*$) denotes that the respective effect is significant at the 5\% (10\%) level, $^\circ$ implies that it is not significant at the 10\% level. Grey font symbolises that the respective variable is not included in the final model. 
$R_\text{av.}^2$ ($\bar{R}_\text{av.}^2$) denotes the average (adjusted) $R^2$ across all 10 imputed data sets.}\\
\toprule
\label{Tab:MIRL_logNAB}
 & \multicolumn{9}{c}{Balanced Panel} & \multicolumn{9}{c}{Unbalanced Panel}\\
 \cmidrule[0.5pt](lr{.75em}){2-10} \cmidrule[0.5pt](lr{.75em}){11-19}

 & \multicolumn{3}{c}{\textbf{BIC}} & \multicolumn{3}{c}{\textbf{AIC}} & \multicolumn{3}{c}{\textbf{C}$_\textbf{p}$} & 
  \multicolumn{3}{c}{\textbf{BIC}} & 
  \multicolumn{3}{c}{\textbf{AIC}} & \multicolumn{3}{c}{\textbf{C}$_\textbf{p}$}\\
  \cmidrule[0.5pt](lr{.75em}){2-4} \cmidrule[0.5pt](lr{.75em}){5-7} \cmidrule[0.5pt](lr{.75em}){8-10} \cmidrule[0.5pt](lr{.75em}){11-13} \cmidrule[0.5pt](lr{.75em}){14-16} \cmidrule[0.5pt](lr{.75em}){17-19}
 \textbf{Variable} & ${\hat{\pi}}$ & $b_{\text{MIRL}}$ & sd$_{b_{\text{MIRL}}}$ & ${\hat{\pi}}$ & ${b_{\text{MIRL}}}$ & sd$_{b_{\text{MIRL}}}$ & ${\hat{\pi}}$ & ${b_{\text{MIRL}}}$ & sd$_{b_{\text{MIRL}}}$ & ${\hat{\pi}}$ & $b_{\text{MIRL}}$ & sd$_{b_{\text{MIRL}}}$ & ${\hat{\pi}}$ & ${b_{\text{MIRL}}}$ & sd$_{b_{\text{MIRL}}}$ & ${\hat{\pi}}$ & ${b_{\text{MIRL}}}$ & sd$_{b_{\text{MIRL}}}$\\
\midrule
\endfirsthead
\caption[]{Summary of the MIRL estimation results based on the balanced and unbalanced panel data for target variable \textbf{\textit{log(NAB)}}. \textit{(continued)}}\\
\toprule
 & \multicolumn{9}{c}{Balanced Panel} & \multicolumn{9}{c}{Unbalanced Panel}\\ \cmidrule[0.5pt](lr{.75em}){2-10} \cmidrule[0.5pt](lr{.75em}){11-19}

   & \multicolumn{3}{c}{\textbf{BIC}} & \multicolumn{3}{c}{\textbf{AIC}} & \multicolumn{3}{c}{\textbf{C}$_\textbf{p}$} & \multicolumn{3}{c}{\textbf{BIC}} & 
  \multicolumn{3}{c}{\textbf{AIC}} & \multicolumn{3}{c}{\textbf{C}$_\textbf{p}$} \\
  \cmidrule[0.5pt](lr{.75em}){2-4} \cmidrule[0.5pt](lr{.75em}){5-7} \cmidrule[0.5pt](lr{.75em}){8-10} \cmidrule[0.5pt](lr{.75em}){11-13} \cmidrule[0.5pt](lr{.75em}){14-16} \cmidrule[0.5pt](lr{.75em}){17-19}
 \textbf{Variable} & ${\hat{\pi}}$ & $b_{\text{MIRL}}$ & sd$_{b_{\text{MIRL}}}$ & ${\hat{\pi}}$ & ${b_{\text{MIRL}}}$ & sd$_{b_{\text{MIRL}}}$ & ${\hat{\pi}}$ & ${b_{\text{MIRL}}}$ & sd$_{b_{\text{MIRL}}}$ & ${\hat{\pi}}$ & $b_{\text{MIRL}}$ & sd$_{b_{\text{MIRL}}}$ & ${\hat{\pi}}$ & ${b_{\text{MIRL}}}$ & sd$_{b_{\text{MIRL}}}$ & ${\hat{\pi}}$ & ${b_{\text{MIRL}}}$ & sd$_{b_{\text{MIRL}}}$\\
\midrule
\endhead

\bottomrule
\multicolumn{19}{r}{(To be continued)}
\endfoot
\bottomrule
\endlastfoot

\multicolumn{19}{l}{\textbf{\textsc{Costs}}}\\
\colorbox{babyblue!70}{Administrative Exp. $/$ Assets} & \textcolor{gray!70}{$0.603$} & \textcolor{red}{\textcolor{gray!70}{$-0.014$}} & \textcolor{gray!70}{$0.023$} & \textcolor{gray!70}{$0.344$} & \textcolor{red}{\textcolor{gray!70}{$-0.016$}} & \textcolor{gray!70}{$0.031$} & \textcolor{gray!70}{$0.382$} & \textcolor{red}{\textcolor{gray!70}{$-0.018$}} & \textcolor{gray!70}{$0.031$} & $0.822$ & \textcolor{red}{$-0.029$} & $0.022$ & $0.618$ & \textcolor{red}{$-0.045$} & $0.050$ & $0.660$ & \textcolor{red}{$-0.043$} & $0.049$\\
\colorbox{babyblue!70}{Cost per Borrower} & \textcolor{gray!70}{$0.251$} & \textcolor{red}{\textcolor{gray!70}{$-0.002$}} & \textcolor{gray!70}{$0.014$} & $0.555$ & $ 0.030$ & $0.105$ & $0.564$ & $ 0.039$ & $0.110$ & $0.883$ & \textcolor{red}{$-0.052$} & $0.035$ & $0.731$ & \textcolor{red}{$-0.067$} & $0.071$ & $0.721$ & \textcolor{red}{$-0.060$} & $0.065$\\
\colorbox{babyblue!70}{Cost per Loan} & $0.997$ & \textcolor{red}{$-0.103$} & $0.033$ & $0.891$ & \textcolor{red}{$-0.132$} & $0.116$ & $0.911$ & \textcolor{red}{$-0.143$} & $0.121$ & \textcolor{gray!70}{$0.406$} & \textcolor{red}{\textcolor{gray!70}{$-0.009$}} & \textcolor{gray!70}{$0.031$} & \textcolor{gray!70}{$0.293$} & \textcolor{gray!70}{$ 0.003^{*}$} & \textcolor{gray!70}{$0.062$} & \textcolor{gray!70}{$0.286$} & \textcolor{red}{\textcolor{gray!70}{$-0.003^\circ$}} & \textcolor{gray!70}{$0.058$}\\
\end{longtable}
\endgroup{}

\begingroup\fontsize{7}{9}\selectfont

\begin{longtable}[t]{p{3.3cm}wc{0.5cm}wc{0.8cm}wc{0.5cm}wc{0.5cm}wc{0.8cm}wc{0.5cm}wc{0.5cm}wc{0.8cm}wc{0.5cm}wc{0.5cm}wc{0.8cm}wc{0.5cm}wc{0.5cm}wc{0.8cm}wc{0.5cm}wc{0.5cm}wc{0.8cm}wc{0.5cm}wc{0.5cm}wc{0.8cm}wc{0.5cm}}
\caption{Summary of the MIRL estimation results on the balanced and unbalanced panel data for target variable \textbf{\textit{--log(ALBG)}}. The table contains all variables that were selected to at least one of the three final models per target variable. sd denotes the empirical standard deviation of the initial estimates the MIRL estimates are computed from, see equation \eqref{eq:b_init}. All effects are significant at the $1\%$ level based on bootstrap confidence intervals, unless otherwise stated: 
$^{**}$ ($^*$) denotes that the respective effect is significant at the 5\% (10\%) level, $^\circ$ implies that it is not significant at the 10\% level. Grey font symbolises that the respective variable is not included in the final model. 
$R_\text{av.}^2$ ($\bar{R}_\text{av.}^2$) denotes the average (adjusted) $R^2$ across all 10 imputed data sets.}\\
\toprule
\label{Tab:MIRL_neglogALBG}
 & \multicolumn{9}{c}{Balanced Panel} & \multicolumn{9}{c}{Unbalanced Panel}\\
 \cmidrule[0.5pt](lr{.75em}){2-10} \cmidrule[0.5pt](lr{.75em}){11-19}

 & 
  \multicolumn{3}{c}{\textbf{BIC}} & 
  \multicolumn{3}{c}{\textbf{AIC}} & \multicolumn{3}{c}{\textbf{C}$_\textbf{p}$} & 
  \multicolumn{3}{c}{\textbf{BIC}} & 
  \multicolumn{3}{c}{\textbf{AIC}} & \multicolumn{3}{c}{\textbf{C}$_\textbf{p}$} \\ \cmidrule[0.5pt](lr{.75em}){2-4} \cmidrule[0.5pt](lr{.75em}){5-7} \cmidrule[0.5pt](lr{.75em}){8-10} \cmidrule[0.5pt](lr{.75em}){11-13} \cmidrule[0.5pt](lr{.75em}){14-16} \cmidrule[0.5pt](lr{.75em}){17-19}
 \textbf{Variable} & ${\hat{\pi}}$ & $b_{\text{MIRL}}$ & sd$_{b_{\text{MIRL}}}$ & ${\hat{\pi}}$ & ${b_{\text{MIRL}}}$ & sd$_{b_{\text{MIRL}}}$ & ${\hat{\pi}}$ & ${b_{\text{MIRL}}}$ & sd$_{b_{\text{MIRL}}}$ & ${\hat{\pi}}$ & $b_{\text{MIRL}}$ & sd$_{b_{\text{MIRL}}}$ & ${\hat{\pi}}$ & ${b_{\text{MIRL}}}$ & sd$_{b_{\text{MIRL}}}$ & ${\hat{\pi}}$ & ${b_{\text{MIRL}}}$ & sd$_{b_{\text{MIRL}}}$\\
\midrule
\endfirsthead
\caption[]{Summary of the MIRL estimation results based on the balanced and unbalanced panel data for target variable \textbf{\textit{--log(ALBG)}}. \textit{(continued)}}\\
\toprule
 & \multicolumn{9}{c}{Balanced Panel} & \multicolumn{9}{c}{Unbalanced Panel}\\ \cmidrule[0.5pt](lr{.75em}){2-10} \cmidrule[0.5pt](lr{.75em}){11-19}

  & \multicolumn{3}{c}{\textbf{BIC}} & 
  \multicolumn{3}{c}{\textbf{AIC}} & \multicolumn{3}{c}{\textbf{C}$_\textbf{p}$} & 
  \multicolumn{3}{c}{\textbf{BIC}} & 
  \multicolumn{3}{c}{\textbf{AIC}} & \multicolumn{3}{c}{\textbf{C}$_\textbf{p}$}\\ \cmidrule[0.5pt](lr{.75em}){2-4} \cmidrule[0.5pt](lr{.75em}){5-7} \cmidrule[0.5pt](lr{.75em}){8-10} \cmidrule[0.5pt](lr{.75em}){11-13} \cmidrule[0.5pt](lr{.75em}){14-16} \cmidrule[0.5pt](lr{.75em}){17-19}
 \textbf{Variable} & ${\hat{\pi}}$ & $b_{\text{MIRL}}$ & sd$_{b_{\text{MIRL}}}$ & ${\hat{\pi}}$ & ${b_{\text{MIRL}}}$ & sd$_{b_{\text{MIRL}}}$ & ${\hat{\pi}}$ & ${b_{\text{MIRL}}}$ & sd$_{b_{\text{MIRL}}}$ & ${\hat{\pi}}$ & $b_{\text{MIRL}}$ & sd$_{b_{\text{MIRL}}}$ & ${\hat{\pi}}$ & ${b_{\text{MIRL}}}$ & sd$_{b_{\text{MIRL}}}$ & ${\hat{\pi}}$ & ${b_{\text{MIRL}}}$ & sd$_{b_{\text{MIRL}}}$\\
\midrule
\endhead

\bottomrule
\multicolumn{19}{r}{(To be continued)}
\endfoot
\bottomrule
\endlastfoot

\multicolumn{19}{l}{\textbf{\textsc{Costs}}}\\
\colorbox{babyblue!70}{Cost per Borrower} & $0.904$ & \textcolor{red}{$-0.089$} & $0.061$ & $0.859$ & \textcolor{red}{$-0.116$} & $0.064$ & $0.830$ & \textcolor{red}{$-0.114$} & $0.067$ & $0.965$ & \textcolor{red}{$-0.124$} & $0.149$ & $0.944$ & \textcolor{red}{$-0.244$} & $0.225$ & $0.952$ & \textcolor{red}{$-0.242$} & $0.218$\\
\colorbox{babyblue!70}{Cost per Loan} & $0.605$ & \textcolor{red}{$-0.025$} & $0.055$ & \textcolor{gray!70}{$0.052$} & \textcolor{red}{\textcolor{gray!70}{$-0.002$}} & \textcolor{gray!70}{$0.032$} & \textcolor{gray!70}{$0.039$} & \textcolor{gray!70}{$ 0.001^\circ$} & \textcolor{gray!70}{$0.032$} & \textcolor{gray!70}{$0.311$} & \textcolor{gray!70}{$ 0.040$} & \textcolor{gray!70}{$0.147$} & \textcolor{gray!70}{$0.797$} & \textcolor{gray!70}{$ 0.138$} & \textcolor{gray!70}{$0.226$} & \textcolor{gray!70}{$0.787$} & \textcolor{gray!70}{$ 0.137$} & \textcolor{gray!70}{$0.220$}\\
\colorbox{babyblue!70}{Fin. Exp. Fund. Liab. $/$ Assets} & $0.639$ & $ 0.050$ & $0.109$ & $0.988$ & $ 0.231$ & $0.091$ & $0.990$ & $ 0.236$ & $0.091$ & \textcolor{gray!70}{$0.370$} & \textcolor{gray!70}{$ 0.028$} & \textcolor{gray!70}{$0.068$} & \textcolor{gray!70}{$0.905$} & \textcolor{gray!70}{$ 0.129$} & \textcolor{gray!70}{$0.082$} & \textcolor{gray!70}{$0.893$} & \textcolor{gray!70}{$ 0.128$} & \textcolor{gray!70}{$0.082$}\\
\colorbox{babyblue!70}{Financial Expense $/$ Assets} & $0.491$ & \textcolor{red}{$-0.045$} & $0.107$ & $0.981$ & \textcolor{red}{$-0.222$} & $0.090$ & $0.982$ & \textcolor{red}{$-0.227$} & $0.092$ & \textcolor{gray!70}{$0.671$} & \textcolor{red}{\textcolor{gray!70}{$-0.045$}} & \textcolor{gray!70}{$0.070$} & \textcolor{gray!70}{$0.898$} & \textcolor{red}{\textcolor{gray!70}{$-0.138$}} & \textcolor{gray!70}{$0.082$} & \textcolor{gray!70}{$0.909$} & \textcolor{red}{\textcolor{gray!70}{$-0.137$}} & \textcolor{gray!70}{$0.082$}\\
\end{longtable}
\endgroup{}

\end{landscape}

\end{appendix}

\pagebreak
\setcounter{page}{1}
\renewcommand*{\thepage}{\arabic{page}}


\title{Supplementary Material for "Model Determination for High-Dimensional Longitudinal Data with Missing Observations: An Application to Microfinance Data"}

\vspace{0.2cm}

\begin{aug}
\author[A]{\fnms{Lotta} \snm{Rüter}\ead[label=e1]{lotta.rueter@kit.edu}} \and
\author[A,B]{\fnms{Melanie} \snm{Schienle}\ead[label=e2]{melanie.schienle@kit.edu}}

\address[A]{Institute of Statistics (STAT), Karlsruhe Institute of Technology (KIT)}
\address[B]{Heidelberg Institute of Theoretical Studies (HITS)}

\vspace{0.4cm}

{\footnotesize \text{Corresponding author: Lotta Rüter, \href{mailto:lotta.rueter@kit.edu}{lotta.rueter@kit.edu}}}

\end{aug}

\vspace{1cm}
\beginsupplement
\section{Details on the Construction of the Panel Data Set} \label{Data:Sec:PreProcSteps} 

We construct the balanced panel used in this work as follows. First, the \textit{Financial Performance Data Set in USD} and the \textit{Social Performance Data Set} from the MIX are merged with an inner join regarding the variables \textit{MFI ID}, \textit{MFI Name}, \textit{Fiscal Year}, \textit{As of Date} and \textit{Period Type}. The resulting data set comprises annual data of 1026 MFIs and observations of 472 variables from 2007 to 2018.
From this data, we select the largest panel that includes as many years as possible. We obtain $w^*=[2009,\;2014]$ as described in Section~\ref{Sec:Data}. The pre-processing is continued with all variables in the merged data set and only those $N_6 = 213$ MFIs and years contained in $w^*$.

Subsequently, all variables with more than 50\% missing observations in the chosen window are dropped. This results in the removal of 148 variables including all variables from the financial performance categories \textit{Deposit Products}, \textit{Digital Delivery Channels}, \textit{Enterprises Financed}, \textit{Job Creation}, \textit{Non-Financial Services}, \textit{Poverty Outreach} and \textit{Products}.
 
The remaining 324 variables are examined in more depth. The main objective of this work requires the data set to be comprehensive in such a way that it contains all potentially meaningful and important variables and no redundant or uninformative data. Moreover, the data must be comparable between MFIs of different sizes. All these aspects are addressed in an appropriate manner.
While some variables are removed, others are transformed or simply kept for the subsequent analysis. All steps undertaken are documented in Table~\ref{Appendix:Tab:variable_transformation}, which also provides definitions of all included variables. Any values deemed implausible (e.g., percentages greater than 1) are excluded. The dummy variables in the social performance data are transformed according to the procedure described below.

\subsection*{Transformation of Social Performance Dummies} 
Much of the social performance data is recorded using multiple dummy variables. As a consequence, many columns of the original data set contain redundant data. The problem can be illustrated with the following example from the data.
Within the category \textit{Client Protection}, several variables regarding debt collection practices are defined as stated in Table~\ref{Tab:Data.Ex:ClientProt} where each row represents one variable.

\begin{table}[htbp]
    \begin{scriptsize}
	\centering
	\caption{All variables concerning debt collection practices in the original data set.}
	\begin{tabular}{llll}\toprule
		Clear Debt Collection Practices & \multicolumn{1}{l}{No} & \multicolumn{1}{r}{} &  \\
		\ditto & \multicolumn{1}{l}{Partially} & \multicolumn{1}{r}{} &  \\
		\ditto & \ditto & \multicolumn{1}{l}{Clear Sanctions for Violations of Debt Collection Practices} & No \\
		\ditto & \ditto & \ditto & Partially \\
		\ditto & \ditto & \ditto & Unknown \\
		\ditto & \ditto & \ditto & Yes \\
		\ditto & \multicolumn{1}{l}{Unknown} & \multicolumn{1}{r}{} &  \\
		\ditto & \multicolumn{1}{l}{Yes} & \multicolumn{1}{r}{} &  \\
		\ditto & \ditto & \multicolumn{1}{l}{Clear Sanctions for Violations of Debt Collection Practices} & No \\
		\ditto & \ditto & \ditto & Partially \\
		\ditto & \ditto & \ditto & Unknown \\
		\ditto & \ditto & \ditto & Yes \\\bottomrule
	\end{tabular}%
	\label{Tab:Data.Ex:ClientProt}%
    \end{scriptsize}
\end{table}%

As a result, the original data set contains 12 different variables that essentially carry the same or at least very similar information. Such variables are reduced to one single dummy variable. In this example, the new variable is defined as

\begin{scriptsize}
\begin{equation} \label{Data:Eq:DummyDef}
	\text{\textit{Clear Debt Collection Practices}} = \begin{cases}
		1, & \text{if \textit{Clear Debt Collection Practices $\triangleright$ Yes}}=1, \\
		0, & \text{if \textit{Clear Debt Collection Practices $\triangleright$ No}}=1,\\
		\textit{NA}, & \text{otherwise}.
	\end{cases}
\end{equation}
\end{scriptsize}

Additional sub-variables, such as those providing details on \textit{Clear Sanctions for Violations of Debt Collection Practices} are deleted. The original data set contains very few observations of variables that are similar (in the way that they contain information on partially fulfilled conditions) to \textit{Clear Debt Collection Practices $\triangleright$ Partially} with value 1. Hence, they are discarded and included as \textit{NA} in the new dummy variables.

{\addtolength\textwidth{0.8cm}

\begin{landscape}
\begingroup\fontsize{7}{9}\selectfont
\renewcommand{\arraystretch}{1.0}

\begin{longtable}[t]{p{0.02cm}p{5.5cm}P{0.5cm}p{3.3cm}p{11.5cm}}
\caption{Details on the construction of the financial and social performance data of the balanced and unbalanced panel data set. Variables are either kept (K), transformed (T) or deleted (D). The operator "\textit{$\triangleright$}" functions as a subcategory indicator in the variable names similar to the original data set. The colours of the new variables indicate the factors they belong to, see Figure \ref{Data:Fig:Piechart-Factors}. The descriptions of the variables are taken from the file \textit{Financial Performance Field Definitions.xlsx} and \textit{Social Performance Field Definitions.xlsx}\footref{Data:Foot:MIX}.}\\
\toprule
\label{Appendix:Tab:variable_transformation}
& \textbf{Original Variable} & \textbf{Action} & \textbf{New Variable} & \textbf{Definition of Reason for Deletion}\\\midrule
\endfirsthead

\caption[]{Details on the construction of the balanced and unbalanced data set. \textit{(continued)}}\\
\toprule
 & \textbf{Original Variable} & \textbf{Action} & \textbf{New Variable} & \textbf{Definition of Reason for Deletion}\\\midrule
\endhead

\bottomrule
\multicolumn{5}{r}{(To be continued)}
\endfoot
\bottomrule
\endlastfoot

\multicolumn{5}{c}{\textbf{\textit{Financial Performance Variables}}}\\
\midrule\midrule

\multicolumn{5}{l}{\textbf{\textit{Balance Sheet}}}\\
& Assets & K & \colorbox{babyblue!20}{Assets} & Total value of resources controlled by the MFI as a result of past events and from which future economic benefits are expected to flow to the MFI; for calculation purposes, assets are the sum of each individual asset account \\
& Average Assets & D &  & Redundant, similar to \textit{Assets} \\
& Borrowings & T & \colorbox{babyblue!20}{Borrowings $/$ Liabilities} &  \textit{Borrowings $/$ Liabilities}; Borrowings denote the principal balance for all funds received through a loan agreement, may include bonds or similar debt securities issued and credit lines \\
& Average Assets & D &  & Redundant, similar to \textit{Assets} \\
& Borrowings & T & \colorbox{babyblue!20}{Borrowings $/$ Liabilities} &  \textit{Borrowings $/$ Liabilities}; Borrowings denote the principal balance for all funds received through a loan agreement, may include bonds or similar debt securities issued and credit lines \\
& Cash and Cash Equivalents & D &  & Redundant: contained in \textit{\% Assets: Non-Earn. Liq. Ass.} \\
\end{longtable}
\endgroup{}
\end{landscape}}

\section{Descriptive Statistics}\,

\begingroup\fontsize{7}{9}\selectfont

\begin{longtable}[H]{p{3.1cm}P{0.9cm}P{0.9cm}P{0.9cm}P{0.9cm}P{0.9cm}P{1cm}P{0.9cm}HP{1.25cm}}
\caption{Descriptive statistics of the balanced panel data used in this work. It contains 1278 observations of 213 MFIs from 2009 to 2014. $Q1$ and $Q3$ denote the first (third) quartile. $\% NA$ gives the resulting proportion of missing values and $N_I$ is the number of implausible values that were manually removed from the analysis.}\\\toprule
\label{Appendix:Tab:DescrStatsBalanced}
\textbf{Variable} & \textbf{Mean} & \textbf{SD} & \textbf{Min.} & \textbf{Q1} & \textbf{Median} & \textbf{Q3} & \textbf{Max.} & $\mathbf{n}_\textbf{imp.}$ & \textbf{\%NA} $(\mathbf{N_I})$ \\
\midrule
\endfirsthead
\caption{Descriptive statistics for the balanced panel data. \textit{(continued)}}\\
\toprule
\textbf{Variable} & \textbf{Mean} & \textbf{SD} & \textbf{Min.} & \textbf{Q1} & \textbf{Median} & \textbf{Q3} & \textbf{Max.} & $\mathbf{n}_\textbf{imp.}$ & \textbf{\%NA} $(\mathbf{N_I})$ \\
\midrule
\endhead

\bottomrule
\multicolumn{10}{r}{(To be continued)}
\endfoot
\bottomrule
\endlastfoot

\multicolumn{10}{l}{\textbf{\textsc{Costs}}}\\
\colorbox{babyblue!70}{Administrative Exp. $/$ Assets} & 0.070 & 0.054 & 0.000 & 0.041 & 0.056 & 0.084 & 0.554 & 0 & 0.042 (0)\\
\colorbox{babyblue!70}{Cost per Borrower} & 215.845 & 237.845 & 5.000 & 93.000 & 164.000 & 271.000 & 4003.000 & 0 & 0.052 (0)\\
\colorbox{babyblue!70}{Cost per Loan} & 198.964 & 206.855 & 5.000 & 89.500 & 160.000 & 254.000 & 4003.000 & 0 & 0.052 (0)\\
\colorbox{babyblue!70}{Fin. Exp. Fund. Liab. $/$ Assets} & 0.054 & 0.033 & -0.370 & 0.034 & 0.052 & 0.070 & 0.287 & 0 & 0.106 (0)\\
\colorbox{babyblue!70}{Financial Expense $/$ Assets} & 0.058 & 0.033 & -0.335 & 0.038 & 0.058 & 0.076 & 0.277 & 0 & 0.038 (0)\\
\colorbox{babyblue!70}{Interest Exp. on Borr. $/$ Borr.} & 0.094 & 0.095 & 0.000 & 0.060 & 0.080 & 0.107 & 1.876 & 0 & 0.063 (0)\\
\colorbox{babyblue!70}{Operating Expense $/$ Assets} & 0.170 & 0.108 & 0.009 & 0.098 & 0.144 & 0.212 & 0.954 & 0 & 0.034 (0)\\
\end{longtable}
\endgroup{}

\begingroup\fontsize{7}{9}\selectfont

\begin{longtable}[H]{p{3.1cm}P{0.9cm}P{0.9cm}P{0.9cm}P{0.9cm}P{0.9cm}P{1cm}P{0.9cm}HP{1.25cm}}
\caption{Descriptive statistics of the unbalanced panel data used for reference in this work. It contains 3846 observations of 1026 MFIs over the years 2007 and 2018. $Q1$ and $Q3$ denote the first (third) quartile. $\% NA$ gives the resulting proportion of missing values and $N_I$ is the number of implausible values that were manually removed from the analysis.}
\\\toprule
\label{Appendix:Tab:DescrStatsUnbalanced}
\textbf{Variable} & \textbf{Mean} & \textbf{SD} & \textbf{Min.} & \textbf{Q1} & \textbf{Median} & \textbf{Q3} & \textbf{Max.} & $\mathbf{n}_\textbf{imp.}$ & \textbf{\%NA} $(\mathbf{N_I})$ \\
\midrule
\endfirsthead
\caption{Descriptive statistics for the unbalanced panel data. \textit{(continued)}}\\
\toprule
\textbf{Variable} & \textbf{Mean} & \textbf{SD} & \textbf{Min.} & \textbf{Q1} & \textbf{Median} & \textbf{Q3} & \textbf{Max.} & $\mathbf{n}_\textbf{imp.}$ & \textbf{\%NA} $(\mathbf{N_I})$ \\
\midrule
\endhead

\bottomrule
\multicolumn{10}{r}{(To be continued)}
\endfoot
\bottomrule
\endlastfoot

\multicolumn{10}{l}{\textbf{\textsc{Costs}}}\\
\colorbox{babyblue!70}{Administrative Exp. $/$ Assets} & 0.077 & 0.075 & -0.050 & 0.037 & 0.056 & 0.090 & 1.179 & 0 & 0.105 (0)\\
\colorbox{babyblue!70}{Cost per Borrower} & 239.261 & 733.423 & -339.000 & 45.000 & 146.000 & 268.000 & 34387.000 & 0 & 0.114 (0)\\
\colorbox{babyblue!70}{Cost per Loan} & 218.831 & 673.751 & -312.000 & 45.000 & 143.000 & 254.000 & 32127.000 & 0 & 0.133 (0)\\
\colorbox{babyblue!70}{Fin. Exp. Fund. Liab. $/$ Assets} & 0.054 & 0.039 & -0.370 & 0.031 & 0.051 & 0.071 & 0.605 & 0 & 0.108 (0)\\
\colorbox{babyblue!70}{Financial Expense $/$ Assets} & 0.058 & 0.039 & -0.335 & 0.033 & 0.056 & 0.079 & 0.878 & 0 & 0.096 (0)\\
\colorbox{babyblue!70}{Interest Exp. on Borr. $/$ Borr.} & 0.131 & 1.562 & -0.121 & 0.056 & 0.080 & 0.113 & 90.911 & 0 & 0.106 (0)\\
\colorbox{babyblue!70}{Operating Expense $/$ Assets} & 0.183 & 0.161 & -0.120 & 0.092 & 0.137 & 0.220 & 3.091 & 0 & 0.093 (0)\\
\end{longtable}
\endgroup{}

{\addtolength\textwidth{0.82cm}
\begin{landscape}
\section{Two-Step Lasso Estimation Results}\,

\begingroup\fontsize{7}{9}\selectfont

\begin{longtable}[t]{p{3.3cm}wc{0.8cm}wc{0.5cm}wc{0.8cm}wc{0.5cm}wc{0.8cm}wc{0.5cm}wc{0.8cm}wc{0.5cm}wc{0.8cm}wc{0.5cm}wc{0.8cm}wc{0.5cm}wc{0.8cm}wc{0.5cm}wc{0.8cm}wc{0.5cm}wc{0.8cm}wc{0.5cm}wc{0.8cm}wc{0.5cm}}
\caption{Summary of two-step lasso estimation results on the balanced and unbalanced panel data for target variable \textbf{\textit{OSS}}. The table contains all variables that were selected to at least one of the three final models per target variable. sd denotes the OLS standard error of the estimate. All effects are significant at the $1\%$ level based on standard confidence intervals under normality, unless otherwise stated: 
$^{**}$ ($^*$) denotes that the respective effect is significant at the 5\% (10\%) level, $^\circ$ implies that it is not significant at the 10\% level. "\textcolor{gray!70}{---}" symbolises that the respective variable is not selected by lasso. 
$R_\text{av.}^2$ ($\bar{R}_\text{av.}^2$) denotes the average (adjusted) $R^2$ across all 10 imputed data sets.}\\
\toprule
\label{Tab:lasso_OSS}
 & \multicolumn{6}{c}{Balanced Panel, Fixed Effects} & \multicolumn{6}{c}{Balanced Panel, Pooled OLS} & \multicolumn{6}{c}{Unbalanced Panel, Pooled OLS}\\
 \cmidrule[0.5pt](lr{.75em}){2-7} \cmidrule[0.5pt](lr{.75em}){8-13} \cmidrule[0.5pt](lr{.75em}){14-19}

  & \multicolumn{2}{c}{\textbf{BIC}} & \multicolumn{2}{c}{\textbf{AIC}} & \multicolumn{2}{c}{\textbf{C}$_\textbf{p}$} & \multicolumn{2}{c}{\textbf{BIC}} & \multicolumn{2}{c}{\textbf{AIC}} & \multicolumn{2}{c}{\textbf{C}$_\textbf{p}$} & \multicolumn{2}{c}{\textbf{BIC}} & \multicolumn{2}{c}{\textbf{AIC}} & \multicolumn{2}{c}{\textbf{C}$_\textbf{p}$}\\ \cmidrule[0.5pt](lr{.75em}){2-3} \cmidrule[0.5pt](lr{.75em}){4-5} \cmidrule[0.5pt](lr{.75em}){6-7} \cmidrule[0.5pt](lr{.75em}){8-9} \cmidrule[0.5pt](lr{.75em}){10-11} \cmidrule[0.5pt](lr{.75em}){12-13} \cmidrule[0.5pt](lr{.75em}){14-15} \cmidrule[0.5pt](lr{.75em}){16-17} \cmidrule[0.5pt](lr{.75em}){18-19}
 \textbf{Variable} & ${b}$ & sd$_{b}$ & ${b}$ & sd$_{b}$ & ${b}$ & sd$_{b}$ & ${b}$ & sd$_{b}$ & ${b}$ & sd$_{b}$ & ${b}$ & sd$_{b}$ & ${b}$ & sd$_{b}$ & ${b}$ & sd$_{b}$ & ${b}$ & sd$_{b}$ \\
\midrule
\endfirsthead
\caption[]{Summary of the two-step lasso estimation results for target variable \textbf{\textit{OSS}}. \textit{(continued)}}\\
\toprule
 & \multicolumn{6}{c}{Balanced Panel, Fixed Effects} & \multicolumn{6}{c}{Balanced Panel, Pooled OLS} & \multicolumn{6}{c}{Unbalanced Panel, Pooled OLS}\\
 \cmidrule[0.5pt](lr{.75em}){2-7} \cmidrule[0.5pt](lr{.75em}){8-13} \cmidrule[0.5pt](lr{.75em}){14-19}

  & \multicolumn{2}{c}{\textbf{BIC}} & \multicolumn{2}{c}{\textbf{AIC}} & \multicolumn{2}{c}{\textbf{C}$_\textbf{p}$} & \multicolumn{2}{c}{\textbf{BIC}} & \multicolumn{2}{c}{\textbf{AIC}} & \multicolumn{2}{c}{\textbf{C}$_\textbf{p}$} & \multicolumn{2}{c}{\textbf{BIC}} & \multicolumn{2}{c}{\textbf{AIC}} & \multicolumn{2}{c}{\textbf{C}$_\textbf{p}$}\\ \cmidrule[0.5pt](lr{.75em}){2-3} \cmidrule[0.5pt](lr{.75em}){4-5} \cmidrule[0.5pt](lr{.75em}){6-7} \cmidrule[0.5pt](lr{.75em}){8-9} \cmidrule[0.5pt](lr{.75em}){10-11} \cmidrule[0.5pt](lr{.75em}){12-13} \cmidrule[0.5pt](lr{.75em}){14-15} \cmidrule[0.5pt](lr{.75em}){16-17} \cmidrule[0.5pt](lr{.75em}){18-19}
 \textbf{Variable} & ${b}$ & sd$_{b}$ & ${b}$ & sd$_{b}$ & ${b}$ & sd$_{b}$ & ${b}$ & sd$_{b}$ & ${b}$ & sd$_{b}$ & ${b}$ & sd$_{b}$ & ${b}$ & sd$_{b}$ & ${b}$ & sd$_{b}$ & ${b}$ & sd$_{b}$ \\
\midrule
\endhead

\bottomrule
\multicolumn{19}{r}{(To be continued)}
\endfoot
\bottomrule
\endlastfoot

\multicolumn{19}{l}{\textbf{\textsc{Costs}}}\\
\colorbox{babyblue!70}{Cost per Loan} & \textcolor{red}{$-0.065$} & $0.018$ & \textcolor{red}{$-0.051^{**}$} & $0.020$ & \textcolor{red}{$-0.052$} & $0.020$ & \textcolor{gray!70}{---} & \textcolor{gray!70}{---} & \textcolor{red}{$-0.055^{**}$} & $0.025$ & \textcolor{red}{$-0.055^{**}$} & $0.025$ & \textcolor{gray!70}{---} & \textcolor{gray!70}{---} & \textcolor{gray!70}{---} & \textcolor{gray!70}{---} & \textcolor{gray!70}{---} & \textcolor{gray!70}{---}\\
\colorbox{babyblue!70}{Financial Expense $/$ Assets} & \textcolor{red}{$-0.091$} & $0.017$ & \textcolor{red}{$-0.104$} & $0.018$ & \textcolor{red}{$-0.102$} & $0.017$ & \textcolor{red}{$-0.060$} & $0.020$ & \textcolor{red}{$-0.072$} & $0.023$ & \textcolor{red}{$-0.072$} & $0.023$ & \textcolor{gray!70}{---} & \textcolor{gray!70}{---} & \textcolor{gray!70}{---} & \textcolor{gray!70}{---} & \textcolor{gray!70}{---} & \textcolor{gray!70}{---}\\
\colorbox{babyblue!70}{Interest Exp. on Borr. $/$ Borr.} & \textcolor{gray!70}{---} & \textcolor{gray!70}{---} & \textcolor{gray!70}{---} & \textcolor{gray!70}{---} & \textcolor{gray!70}{---} & \textcolor{gray!70}{---} & \textcolor{gray!70}{---} & \textcolor{gray!70}{---} & \textcolor{red}{$-0.009^\circ$} & $0.018$ & \textcolor{red}{$-0.009^\circ$} & $0.018$ & \textcolor{gray!70}{---} & \textcolor{gray!70}{---} & \textcolor{gray!70}{---} & \textcolor{gray!70}{---} & \textcolor{gray!70}{---} & \textcolor{gray!70}{---}\\
\colorbox{babyblue!70}{Operating Expense $/$ Assets} & \textcolor{gray!70}{---} & \textcolor{gray!70}{---} & \textcolor{gray!70}{---} & \textcolor{gray!70}{---} & \textcolor{gray!70}{---} & \textcolor{gray!70}{---} & $ 0.003^\circ$ & $0.047$ & \textcolor{gray!70}{---} & \textcolor{gray!70}{---} & \textcolor{gray!70}{---} & \textcolor{gray!70}{---} & \textcolor{gray!70}{---} & \textcolor{gray!70}{---} & \textcolor{gray!70}{---} & \textcolor{gray!70}{---} & \textcolor{gray!70}{---} & \textcolor{gray!70}{---}\\
\end{longtable}
\endgroup{}

\begingroup\fontsize{7}{9}\selectfont

\begin{longtable}[t]{p{3.3cm}wc{0.8cm}wc{0.5cm}wc{0.8cm}wc{0.5cm}wc{0.8cm}wc{0.5cm}wc{0.8cm}wc{0.5cm}wc{0.8cm}wc{0.5cm}wc{0.8cm}wc{0.5cm}wc{0.8cm}wc{0.5cm}wc{0.8cm}wc{0.5cm}wc{0.8cm}wc{0.5cm}wc{0.8cm}wc{0.5cm}}
\caption{Summary of two-step lasso estimation results on the balanced and unbalanced panel data for target variable \textbf{\textit{log(NAB)}}. The table contains all variables that were selected to at least one of the three final models per target variable. sd denotes the OLS standard error of the estimate. All effects are significant at the $1\%$ level based on standard confidence intervals under normality, unless otherwise stated: 
$^{**}$ ($^*$) denotes that the respective effect is significant at the 5\% (10\%) level, $^\circ$ implies that it is not significant at the 10\% level. "\textcolor{gray!70}{---}" symbolises that the respective variable is not selected by lasso. 
$R_\text{av.}^2$ ($\bar{R}_\text{av.}^2$) denotes the average (adjusted) $R^2$ across all 10 imputed data sets.}\\
\toprule
\label{Tab:lasso_logNAB}
 & \multicolumn{6}{c}{Balanced Panel, Fixed Effects} & \multicolumn{6}{c}{Balanced Panel, Pooled OLS} & \multicolumn{6}{c}{Unbalanced Panel, Pooled OLS}\\
 \cmidrule[0.5pt](lr{.75em}){2-7} \cmidrule[0.5pt](lr{.75em}){8-13} \cmidrule[0.5pt](lr{.75em}){14-19}

  & \multicolumn{2}{c}{\textbf{BIC}} & \multicolumn{2}{c}{\textbf{AIC}} & \multicolumn{2}{c}{\textbf{C}$_\textbf{p}$} & \multicolumn{2}{c}{\textbf{BIC}} & \multicolumn{2}{c}{\textbf{AIC}} & \multicolumn{2}{c}{\textbf{C}$_\textbf{p}$} & \multicolumn{2}{c}{\textbf{BIC}} & \multicolumn{2}{c}{\textbf{AIC}} & \multicolumn{2}{c}{\textbf{C}$_\textbf{p}$}\\ \cmidrule[0.5pt](lr{.75em}){2-3} \cmidrule[0.5pt](lr{.75em}){4-5} \cmidrule[0.5pt](lr{.75em}){6-7} \cmidrule[0.5pt](lr{.75em}){8-9} \cmidrule[0.5pt](lr{.75em}){10-11} \cmidrule[0.5pt](lr{.75em}){12-13} \cmidrule[0.5pt](lr{.75em}){14-15} \cmidrule[0.5pt](lr{.75em}){16-17} \cmidrule[0.5pt](lr{.75em}){18-19}
 \textbf{Variable} & ${b}$ & sd$_{b}$ & ${b}$ & sd$_{b}$ & ${b}$ & sd$_{b}$ & ${b}$ & sd$_{b}$ & ${b}$ & sd$_{b}$ & ${b}$ & sd$_{b}$ & ${b}$ & sd$_{b}$ & ${b}$ & sd$_{b}$ & ${b}$ & sd$_{b}$ \\
\midrule
\endfirsthead
\caption[]{Summary of the two-step lasso estimation results for target variable \textbf{\textit{log(NAB)}}. \textit{(continued)}}\\
\toprule
 & \multicolumn{6}{c}{Balanced Panel, Fixed Effects} & \multicolumn{6}{c}{Balanced Panel, Pooled OLS} & \multicolumn{6}{c}{Unbalanced Panel, Pooled OLS}\\
 \cmidrule[0.5pt](lr{.75em}){2-7} \cmidrule[0.5pt](lr{.75em}){8-13} \cmidrule[0.5pt](lr{.75em}){14-19}

  & \multicolumn{2}{c}{\textbf{BIC}} & \multicolumn{2}{c}{\textbf{AIC}} & \multicolumn{2}{c}{\textbf{C}$_\textbf{p}$} & \multicolumn{2}{c}{\textbf{BIC}} & \multicolumn{2}{c}{\textbf{AIC}} & \multicolumn{2}{c}{\textbf{C}$_\textbf{p}$} & \multicolumn{2}{c}{\textbf{BIC}} & \multicolumn{2}{c}{\textbf{AIC}} & \multicolumn{2}{c}{\textbf{C}$_\textbf{p}$}\\ \cmidrule[0.5pt](lr{.75em}){2-3} \cmidrule[0.5pt](lr{.75em}){4-5} \cmidrule[0.5pt](lr{.75em}){6-7} \cmidrule[0.5pt](lr{.75em}){8-9} \cmidrule[0.5pt](lr{.75em}){10-11} \cmidrule[0.5pt](lr{.75em}){12-13} \cmidrule[0.5pt](lr{.75em}){14-15} \cmidrule[0.5pt](lr{.75em}){16-17} \cmidrule[0.5pt](lr{.75em}){18-19}
 \textbf{Variable} & ${b}$ & sd$_{b}$ & ${b}$ & sd$_{b}$ & ${b}$ & sd$_{b}$ & ${b}$ & sd$_{b}$ & ${b}$ & sd$_{b}$ & ${b}$ & sd$_{b}$ & ${b}$ & sd$_{b}$ & ${b}$ & sd$_{b}$ & ${b}$ & sd$_{b}$ \\
\midrule
\endhead

\bottomrule
\multicolumn{19}{r}{(To be continued)}
\endfoot
\bottomrule
\endlastfoot

\multicolumn{19}{l}{\textbf{\textsc{Costs}}}\\
\colorbox{babyblue!70}{Administrative Exp. $/$ Assets} & \textcolor{gray!70}{---} & \textcolor{gray!70}{---} & \textcolor{gray!70}{---} & \textcolor{gray!70}{---} & \textcolor{gray!70}{---} & \textcolor{gray!70}{---} & \textcolor{gray!70}{---} & \textcolor{gray!70}{---} & \textcolor{red}{$-0.065$} & $0.021$ & \textcolor{red}{$-0.063$} & $0.021$ & \textcolor{red}{$-0.032$} & $0.012$ & \textcolor{red}{$-0.061$} & $0.020$ & \textcolor{red}{$-0.061$} & $0.020$\\
\colorbox{babyblue!70}{Cost per Borrower} & \textcolor{red}{$-0.050$} & $0.017$ & \textcolor{red}{$-0.042^{**}$} & $0.017$ & \textcolor{red}{$-0.042^{**}$} & $0.017$ & \textcolor{gray!70}{---} & \textcolor{gray!70}{---} & \textcolor{gray!70}{---} & \textcolor{gray!70}{---} & \textcolor{gray!70}{---} & \textcolor{gray!70}{---} & \textcolor{red}{$-0.039$} & $0.010$ & \textcolor{red}{$-0.035$} & $0.010$ & \textcolor{red}{$-0.035$} & $0.010$\\
\colorbox{babyblue!70}{Cost per Loan} & \textcolor{gray!70}{---} & \textcolor{gray!70}{---} & \textcolor{gray!70}{---} & \textcolor{gray!70}{---} & \textcolor{gray!70}{---} & \textcolor{gray!70}{---} & \textcolor{red}{$-0.131$} & $0.018$ & \textcolor{red}{$-0.113$} & $0.019$ & \textcolor{red}{$-0.113$} & $0.019$ & \textcolor{gray!70}{---} & \textcolor{gray!70}{---} & \textcolor{gray!70}{---} & \textcolor{gray!70}{---} & \textcolor{gray!70}{---} & \textcolor{gray!70}{---}\\
\colorbox{babyblue!70}{Fin. Exp. Fund. Liab. $/$ Assets} & \textcolor{gray!70}{---} & \textcolor{gray!70}{---} & \textcolor{gray!70}{---} & \textcolor{gray!70}{---} & \textcolor{gray!70}{---} & \textcolor{gray!70}{---} & \textcolor{gray!70}{---} & \textcolor{gray!70}{---} & \textcolor{gray!70}{---} & \textcolor{gray!70}{---} & \textcolor{gray!70}{---} & \textcolor{gray!70}{---} & \textcolor{gray!70}{---} & \textcolor{gray!70}{---} & \textcolor{red}{$-0.083$} & $0.021$ & \textcolor{red}{$-0.083$} & $0.021$\\
\end{longtable}%
\endgroup{}
\end{landscape}
}

\end{document}